\documentclass[prl,singlecolumn,preprint,superscriptaddress]{revtex4-1}
\usepackage{bm}
\usepackage[colorlinks=true,linkcolor=blue,citecolor=blue]{hyperref}
\usepackage{times}
\usepackage{amsmath}
\usepackage{amssymb}
\usepackage{amsthm}
\usepackage{amsfonts}
\usepackage{enumerate}
\usepackage{latexsym}
\usepackage{ifpdf}
\usepackage{natbib}
\usepackage{psfrag}
\newcommand{\beq}{\begin{equation}}
\newcommand{\eeq}{\end{equation}}
\usepackage{graphicx}
\usepackage{makeidx}
\usepackage{color}
\usepackage{array}
\usepackage{multirow}
\usepackage{siunitx}
\usepackage{}
\usepackage{physics}

\begin{document}

\title{Oxygen Vacancy-Induced  Topological  Hall effect in a Nonmagnetic Band Insulator}

\author {Shashank Kumar Ojha}
\affiliation  {Department of Physics, Indian Institute of Science, Bengaluru 560012, India}
\author {Sanat Kumar Gogoi}
\affiliation  {Department of Physics, Indian Institute of Science, Bengaluru 560012, India}
\author{Manju Mishra Patidar}
\affiliation {UGC-DAE CSR, University Campus - Khandawa Road, Indore 452 017, India}
\author {Ranjan Kumar Patel}
\affiliation  {Department of Physics, Indian Institute of Science, Bengaluru 560012, India}
\author {Prithwijit Mandal}
\affiliation  {Department of Physics, Indian Institute of Science, Bengaluru  560012, India}
\author {Siddharth Kumar}
\affiliation  {Department of Physics, Indian Institute of Science, Bengaluru  560012, India}
\author {R. Venkatesh}
\affiliation  {UGC-DAE CSR, University Campus - Khandawa Road, Indore 452 017, India}
\author {V. Ganesan}
\affiliation  {UGC-DAE CSR, University Campus - Khandawa Road, Indore 452 017, India}
\author {Manish Jain}
\affiliation  {Department of Physics, Indian Institute of Science, Bengaluru 560012, India}
\author {Srimanta Middey}
\email{smiddey@iisc.ac.in}
\affiliation  {Department of Physics, Indian Institute of Science, Bengaluru 560012, India}

\begin{abstract}
The discovery of skyrmions has sparked tremendous interests about topologically nontrivial spin textures in recent times. The signature of noncoplanar nature of magnetic moments can be observed as topological Hall effect (THE) in electrical measurement. Realization of such nontrivial spin textures in new materials and through new routes is an ongoing endeavour due to their huge potential for  future ultra-dense low-power memory applications. In this work, we report oxygen vacancy (OV) induced THE and anomalous Hall effect (AHE) in a 5$d^0$ system KTaO$_3$. The observation of weak antilocalization behavior and THE in the same temperature range strongly implies the crucial role of spin-orbit coupling (SOC) behind the origin of THE. Ab initio calculations reveal the formation of the magnetic moment on Ta atoms around the OV and Rashba-type spin texturing of conduction electrons. In the presence of Rashba SOC, the local moments around vacancy can form bound magnetic polarons (BMP) with noncollinear spin texture, resulting THE. Scaling analysis between transverse and longitudinal resistance establishes skew scattering driven AHE in present case. Our study opens a route to realize topological phenomena through defect engineering.
\end{abstract}

\maketitle

A current-carrying ferromagnetic conductor placed in a perpendicular magnetic field exhibits AHE~\cite{nagaosa:2010p1539}. The additional Hall voltage is empirically proportional to the magnetization of the sample, leading to non-zero Hall voltage even when the magnetic field is reduced to zero. While, the mechanism of AHE is highly debated, various proposed mechanisms can be broadly classified into intrinsic (Berry curvature in momentum space) or extrinsic (skew or side jump scattering) types~\cite{nagaosa:2010p1539}. Materials with chiral spin texture having characteristic non-zero scalar spin chirality [$\chi_{ijk}= \textbf{S}_i.(\textbf{S}_j \cross \textbf{S}_k)$ where $\textbf{S}_{i,j,k}$ denote localized magnetic moments] experience additional  fictitious magnetic field due to the real space Berry phase.
  This leads to the observation of THE in various systems including helimagnet~\cite{neubauer:2009p186602,kanazawa:2011p156603}, noncollinear antiferromagnet~\cite{surgers:2014p3400},  frustrated triangular lattice~\cite{kurumaji:2019p914}, magnetic topological insulator~\cite{liu:2017p176809} etc. Since Dzyaloshinskii-Moriya interaction (the key ingredient for stabilizing noncoplanar spin texture) can be introduced by a combined effect of broken inversion symmetry and SOC, THE has been also engineered in various artificial structures including oxide heterostructures~\cite{matsuno:2016p1600304,qin:2019p18007008,vistoli:2019p67,Takahashi:2018p7880,Nakamura:2018p074704}, heterostructure composed of a magnetic insulator and metal~\cite{shao:2019p182}, magnetic multilayers~\cite{raju:2019p1} etc. Interestingly, the recent report of THE above the ferromagnetic $T_c$ of SrRuO$_3$ and V doped Sb$_2$Te$_3$ thin films implies that the presence of non-zero $\chi_{ijk}$ even in a short-range length scale can also generate THE~\cite{Wang:2019p1054}.

Ever since it's discovery, the observation of THE has been unintentionally restricted to only those materials which possess localized magnetic moments.  Interestingly, magnetic moments can be introduced in a nonmagnetic system like SrTiO$_3$ through OV creation~\cite{Coey:2016p485001,Coey:2019p652,Rice:2014p481,Chungwei:2013p217601,Lopez:2015p115112}. Owing to the  presence of Rashba-type SOC~\cite{Rashba:1984p6039}, the nature of  spin arrangements near the surface region can be significantly different compared to that of the bulk, and can lead to emergent phenomena. Due to the possibility of achieving  new set of electronic, magnetic and topological phases as a result  of strong intrinsic SOC, electron doping in an analogous 5$d^0$ compound  KTaO$_3$ (KTO)- based systems are being investigated in recent years ~\cite{nakamura:2009p121308,king:2012p117602,zhang:2018p116803,wadehra:2019planar}.  Bulk KTO is a nonmagnetic band insulator with cubic structure and becomes metallic and superconducting upon electron doping~\cite{ueno:2011p408,harashima:2013p085102}. Interestingly,    unconventional spin-orbital texture/Rashba type spin texture of conduction electrons  have been also claimed in STO and KTO based systems ~\cite{He:2018p266802,Vaz:2019p1187,Altmeyer:2016p157203,Bruno:2019p1800860}. Presence of such conduction electrons with strong SOC can lead  to the formation of  noncollinear BMP of defect-induced localized magnetic moments~\cite{coey:2005p173,denisov:2018p162409}, which would result THE. In spite of such exciting possibility, to date there is no report of either non-trivial spin texturing of local moments or THE in such cases.

  In this work, we report about the magnetotransport behavior of oxygen deficient KTO.  We found that the longitudinal resistance ($R_{xx}$) of such metallic KTaO$_{3-\delta}$ not only exhibits Shubnikov-de Haas  (SdH) oscillation due to  high mobility, but also shows  weak antilocalization (WAL) behavior. Moreover, the transverse resistance ($R_{xy}$) demonstrates signature of both AHE and THE. Interestingly, THE vanishes around 35 K where WAL also becomes insignificant, clearly demonstrating strong interconnection between SOC and observed THE in present case. Scaling analysis further demonstrates skew scattering mechanism as the origin of observed AHE.  Presence of Rashba type spin texturing of conduction electrons is further demonstrated by \textit{ab initio} calculations, strongly suggesting the existence of defect induced noncollinear BMPs behind the observed THE in present case.

\begin{figure*}
	\centering{
		\includegraphics[scale=0.5]{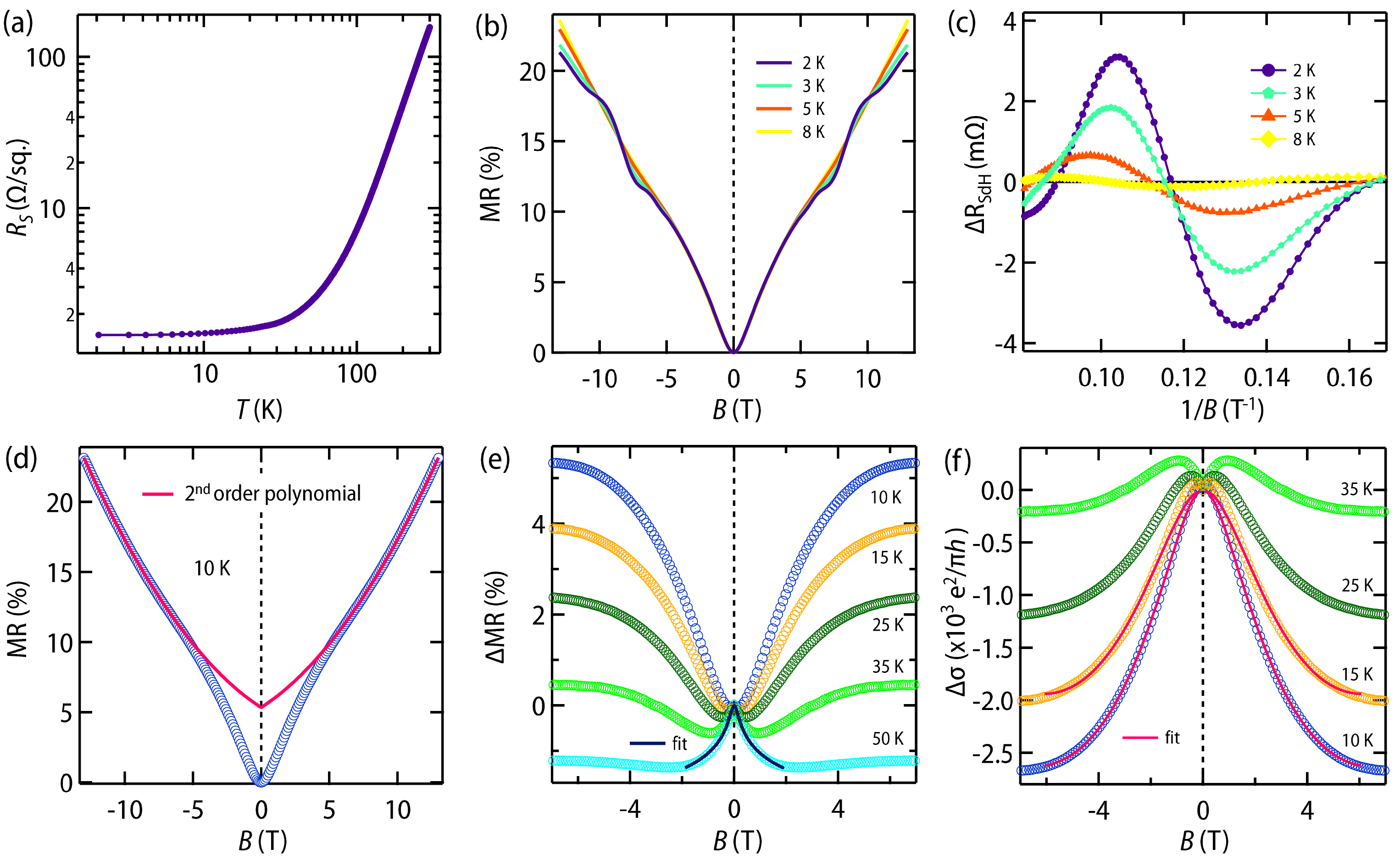}
		\caption{SdH oscillations and weak antilocalization. a) Temperature dependence of sheet resistance $R_S$ of KTO with OV, showing metallic behavior. b)  Symmetrized MR from 2 to 8 K (see Figure S3, Supporting Information for systematic evolution of MR from 2 to 300 K). Magnetic field was applied along [001] crystallographic axis. c) Amplitude of SdH oscillations $\Delta$R${_\text{SdH}}$, obtained by subtracting 2$^\text{nd}$ order polynomial from raw data, at different temperatures. d) MR at 10 K. To isolate contribution from WAL, high field MR was fitted with 2$^\text{nd}$ order polynomial. e) Extracted $\Delta$MR at different fixed temperatures.  We could not isolate WAL below 10 K due to presence of SdH oscillations at high field. $\Delta$MR at 50 K has been fitted with the modified Khosla-Fisher equation~\cite{Khosla:1970p4084} given by MR= -$a$$^2$ln(1+$b^2$$B$$^2$) where $a$ and $b$ are fitting parameters. f) Extracted sheet conductance difference $\Delta$$\sigma$ ($B$) in units of $e^2$/$\pi$$h$ for different $T$, -$e$ is the electrons charge and $h$ is the Planck's constant. Solid  curve shows the fitting of WAL by ILP theory without considering linear Rashba term.} \label{fig:1}}
\end{figure*}

OVs in pristine single crystalline KTO have been incorporated by thermal annealing within an evacuated sealed quartz tube (see experimental section and  Figure S1, Supporting Information).
 As evident from {\bf Figure} \ref{fig:1}a, creation of OVs turns pristine insulating KTO into a metal (throughout this article we have presented results of this sample, referred to as Sample A. Supporting Information contains similar set of data and analysis for another Sample B).  Perpendicular magnetoresistance MR ($B$) (=$\frac{R_{xx}(B)-R_{xx}(B=0)}{R_{xx}(B=0)}$$\times$100\%,  $B$ is external magnetic field)  exhibits SdH oscillations  ({\bf Figure} \ref{fig:1}b) for magnetic field $B \ge$  5 T and $T\leq$ 8 K over a background of positive MR. The background contribution has been modeled by a 2$^\text{nd}$ order polynomial (linear term arises due to presence of inhomogeneities~\cite{Hu:2008p697} and $B^2$ term corresponds to classical Lorentz force).  SdH oscillations are very prominent after the background subtraction ({\bf Figure} \ref{fig:1}c). Further analysis (Section B, Supporting Information) of SdH oscillations provides a carrier concentration of $n_v$=${5.7\times10^{17}{cm^{-3}}}$ and an effective mass of $m^{\star}$=(0.56$\pm$0.02)$m_{e}$ ($m_{e}$ is the bare electron mass). Using the carrier density obtained from Hall measurements and SdH oscillations, the thickness of electron gas is estimated to be several microns. This implies  the present system is 3-dimensional in nature~\cite{herranz:2007p216803,harashima:2013p085102}.

At 10 K,  MR deviates  from a combination of linear $B$ and $B^2$ terms below 6 T, which resembles WAL behavior ({\bf Figure}~\ref{fig:1}d). This WAL is further prominent in {\bf Figure} \ref{fig:1}e, where we have plotted the  additional component of MR ($\Delta$MR) after subtracting linear $B$ and $B^2$ dependent terms. The WAL behavior persists upto 35 K. Additionally, a cusp like feature appears systematically near $B$=0 and  becomes very prominent at 50 K. This behavior can be accounted by scattering of conduction electron with the localized magnetic moments (also see fitting by Khosla-Fisher equation in {\bf Figure} \ref{fig:1}e~\cite{Khosla:1970p4084}. In order to further analyze the WAL behavior, we have also plotted  the sheet conductance difference ($\Delta$$\sigma$ ($B$)=$\sigma$($B$)-$\sigma$(0) where $\sigma$=1/$R$$_S$) in {\bf Figure}~\ref{fig:1}f. Since KTO is known to host strong Rashba SOC~\cite{nakamura:2009p121308}, these WAL features have been analyzed in-terms of ILP (Iordanskii, Lyanda-Geller, and Pikus) theory, which includes linear and cubic Rashba term to describe WAL~\cite{iordanskii:1994p199}. However, better fits were obtained by retaining only cubic Rashba term similar to previous report on KTO~\cite{nakamura:2009p121308} and STO~\cite{Nakamura:2012p206601} based two-dimensional electron gas. In the absence of linear Rashba term, the correction in $\sigma$ is given by
\begin{eqnarray}
\Delta\sigma (B)&=& N\frac{e^2}{\pi h}\Bigg[\Psi(\frac{1}{2}+\frac{B_\phi}{B}+\frac{B_{SO}}{B}) -\frac{1}{2}\Psi(\frac{1}{2}+\frac{B_\phi}{B}) \nonumber
\\ & &+\frac{1}{2}\Psi(\frac{1}{2}+\frac{B_\phi}{B}+\text{2}\frac{B_{SO}}{B})-\text{ln}\frac{B_\phi+B_{SO}}{B} \nonumber
\\ & & -\frac{1}{2}\text{ln}\frac{B_\phi+\text{2}B_{SO}}{B}  +\frac{1}{2}\text{ln}\frac{B_\phi}{B}\Bigg] \label{eq:2}
\end{eqnarray}
where $\Psi$ is the digamma function, $B_\phi$=$\frac{\hbar}{\text{4e}l_{\phi}^2}$ ($l_{\phi}$ is the phase coherence length), $B_{SO}$=$\frac{\hbar}{\text{4e}l_{SO}^2}$ ($l_{SO}$ is the spin-precession length) and $N$ is a scaling factor. As shown in {\bf Figure} \ref{fig:1}f, ILP theory gives an excellent fit to our extracted data, strongly implying dominant role of Rashba SOC in the magneto-transport behavior of oxygen deficient KTO below 35 K. We obtain $l_{SO}$ $\approx$ 10 nm and $l_{\phi}$ $\approx$ 16 nm from the fitting. The value of $l_{SO}$ is in good agreement with previous literature~\cite{nakamura:2009p121308} whereas $l_{\phi}$ is one order of magnitude less. Since WAL is physically meaningful only when the thickness of the sample is less than $l_{\phi}$~\cite{Bergmann:1984p1},  we infer that  WAL features emerge only from  few nm layers near the surface though the OV induced conducting region is extended deep (micron scale) inside the substrate.

\begin{figure*}
	\centering{
		{~}\hspace*{-0.5cm}
		\includegraphics[scale=.58]{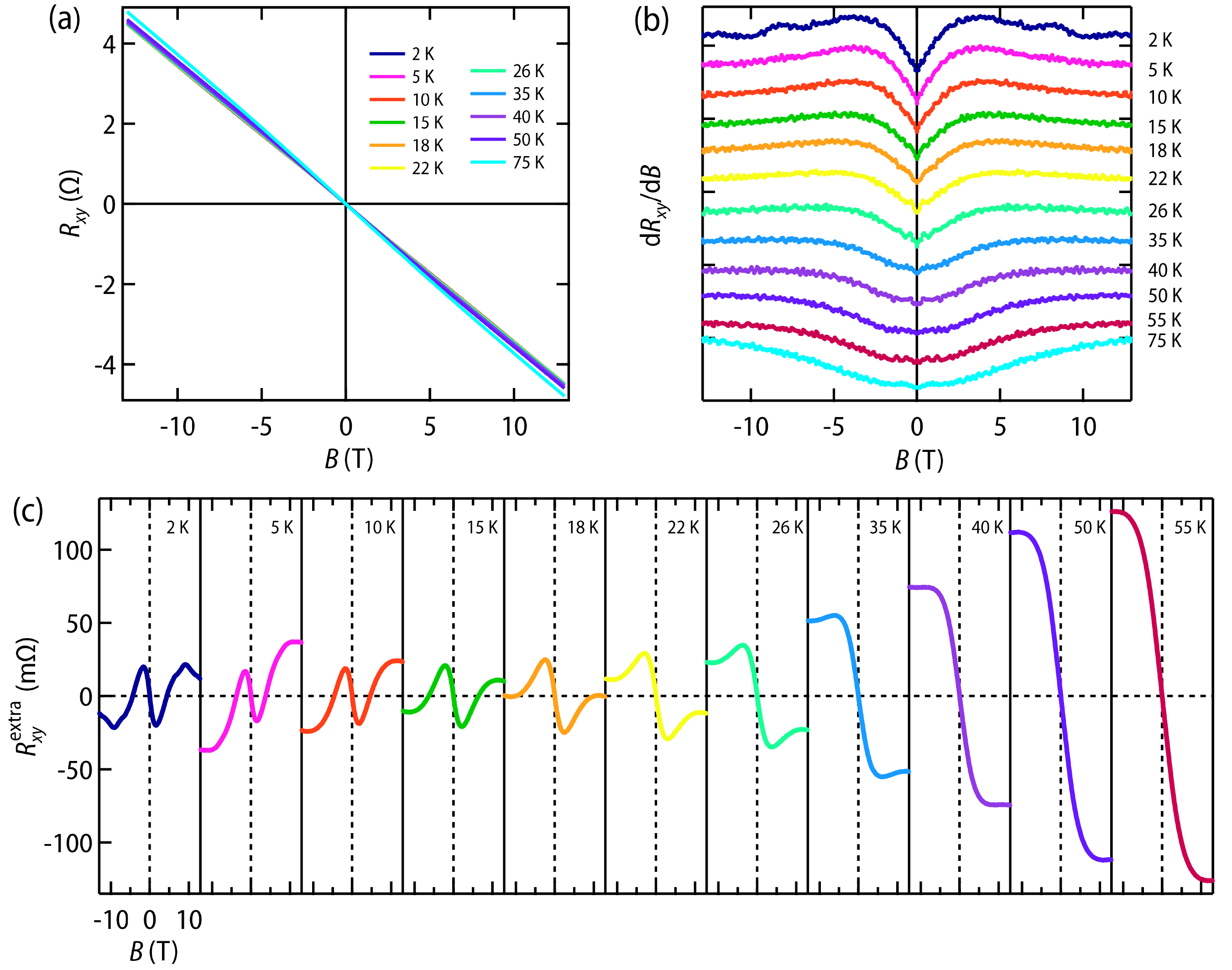}
		\caption{ Non-linearity in Hall resistance $R_{xy}$ and ${R^\text{extra}_{xy}}$ extraction. a)  Magnetic field dependent anti-symmetrized Hall resistance $R$$_{xy}$ measured at different temperatures. b) Derivative of $R$$_{xy}$ w.r.t applied magnetic field, all data have been shifted vertically for visual clarity. c) ${R^\text{extra}_{xy}}$ as a function of magnetic field $B$  at various temperatures from 2 to 55 K. The anomaly at high field at 2 K is due to presence of  oscillations in $\frac{dR_{xy}}{dB}$.  OHE can not be subtracted for $T$ above 55 K as $\frac{dR_{xy}}{dB}$ does not become field independent even at high fields.} \label{fig:2}}
\end{figure*}

Since OVs act as electron donor, the metallic behavior in present case should be $n$-type, which is confirmed by the variation of Hall resistance $R_{xy}$ as a function of $B$ ({\bf Figure} \ref{fig:2}a). However, unlike to the expected field independent behavior of $\frac{dR_{xy}}{dB}$ ($\propto\frac{1}{ne}$, $n$ is carrier concentration) for a simple one-band metal, it shows strong non-linearity in the derivative in the form of a dip at low field ({\bf Figure}~\ref{fig:2}b), strongly suggesting presence of additional effects in the oxygen deficient KTO sample. With increase in $T$, the non-linearity spreads at even higher field. Such non-linearity of ${R_{xy}}$ has been observed before for many  systems including complex oxide interfaces and has been widely attributed to the either  multiband transport or the presence of AHE~\cite{gunkel:2016p031035,takahashi:2009p057204,stornaiuolo:2016p278}. We could not capture this non-linearity  in present case using a two-band model (Figure S4, Supporting Information). Furthermore, at high $B$,  $\frac{dR_{xy}}{dB}$  becomes field-independent and has similar value at all temperatures (Figure S5, Supporting Information),  strongly implying that it is effectively one band transport. We also note that no hysteresis was observed in both MR and Hall measurement.

In presence of AHE and THE, one can write the total Hall resistance $R$$_{xy}$ as
\begin{equation}
R_{xy}(B) = {R^\text{OHE}_{xy}(B)} + {R^\text{extra}_{xy}(B)}  \label{eq:2}
\end{equation}
where ${R^\text{OHE}_{xy}}$ corresponds to the contribution from ordinary Hall effect and  ${R^\text{extra}_{xy}(B)}$  represents  the contribution from AHE and THE.  To estimate ${R^\text{OHE}_{xy}}$, slope of $R$$_{xy}$ (slope was extracted by linear fitting of $R_{xy}$ in the field range where derivative of $R$$_{xy}$ becomes constant) was multiplied with the applied magnetic field. This contribution was then subtracted from $R_{xy}$ to obtain ${R^\text{extra}_{xy}}$.  One such set of extracted ${R^\text{extra}_{xy}(B)}$ are shown in the {\bf Figure}~\ref{fig:2}c.  This extra component of $R_{xy}$ for $T>$ 35 K is a signature of pure AHE~\cite{takahashi:2009p057204,stornaiuolo:2016p278}. Most importantly,  an additional hump-like feature appears below 35 K, which establishes the presence of THE~\cite{Wang:2019p1054}. Interestingly, AHE component of $R_{xy}$ (${R^\text{AHE}_{xy}}$) also changes it's sign around 18 K. Similar sign change has been also observed in SrRuO$_3$~\cite{Wang:2019p1054}.

\begin{figure*}
	\centering{
		\hspace*{-0.3cm}
		\includegraphics[scale=.62]{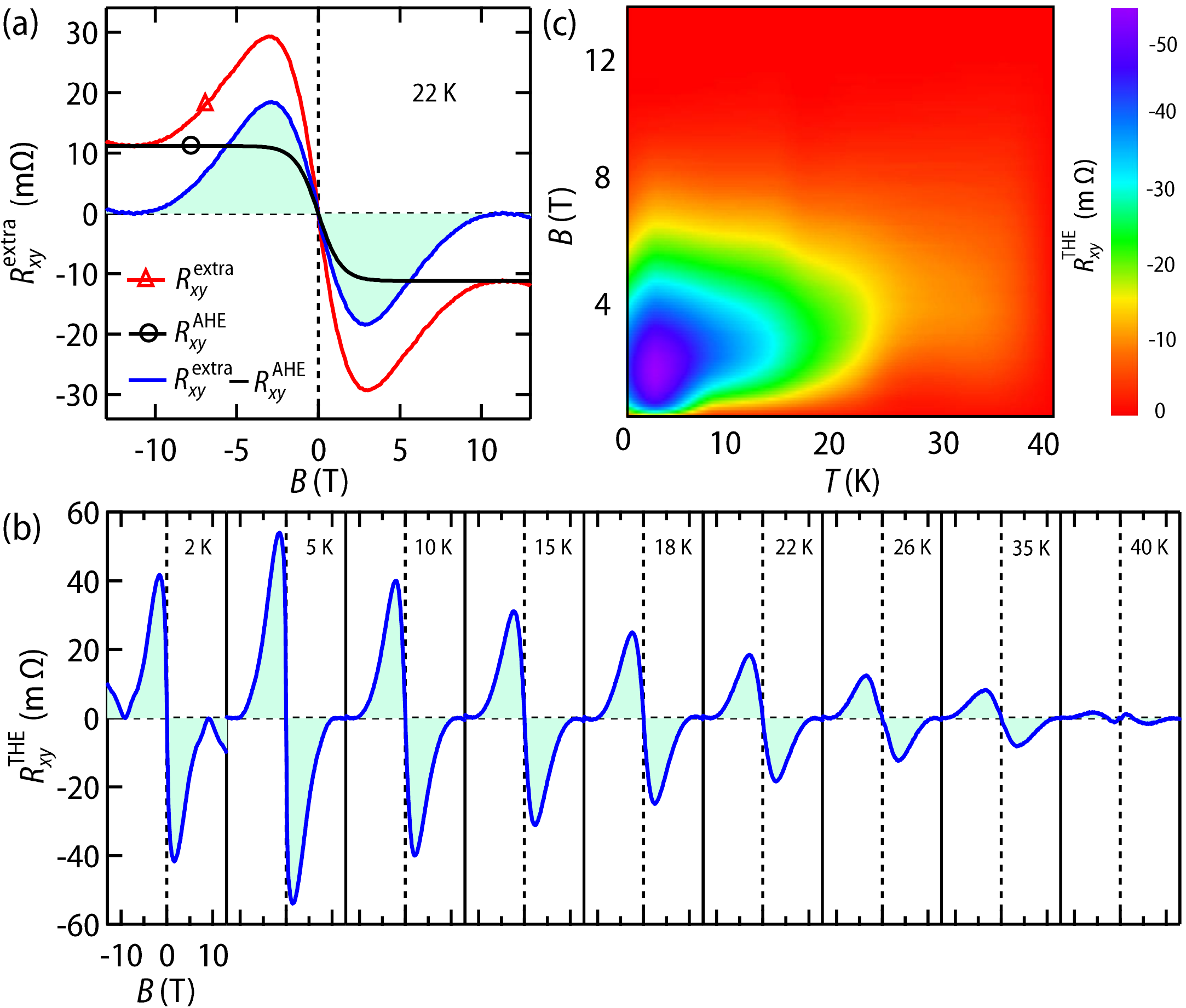}
		\caption{ Extraction of topological Hall resistance ${R^\text{THE}_{xy}}$. a) Fitting of ${R^\text{extra}_{xy}}$ at 22 K with a Brillouin function for estimation of  ${R^\text{AHE}_{xy}}$. b) Extracted ${R^\text{THE}_{xy}}$ as a function of magnetic field $B$ at various temperature from 2 to 40 K. There is anomaly in ${R^\text{THE}_{xy}}$ at high field at 2 K due to presence of  oscillations in $R_{xy}^\text{extra}$ (see Figure \ref{fig:2}b). c) Color map of ${R^\text{THE}_{xy}}$ in the $B$-$T$ plane.} \label{fig:3}}
\end{figure*}

To extract the temperature evolution of the topological Hall resistance ${R^\text{THE}_{xy}}$, we have further subtracted AHE contribution at each temperature from  ${R^\text{extra}_{xy}(B)}$. Since AHE is proportional to magnetization of the sample, we have simulated the contribution of ${R^\text{AHE}_{xy}}$ at each temperature by Brillouin function (see Section G, Supporting Information) which describes the magnetization of a quantum paramagnetic gas at finite temperature in an external magnetic field ({\bf Figure}~\ref{fig:3}a). {\bf Figure}~\ref{fig:3}b shows the extracted ${R^\text{THE}_{xy}}$ from 2 K to 40 K.
As clearly evident, there is a systematic evolution
of hump-like feature from 2 to 40 K. This is the signature of THE and occurs in materials with topologically protected noncoplanar spin texture like skyrmions~\cite{neubauer:2009p186602}. Such THE-like features can also emerge from either two-channel AHE~\cite{Kan:2018p180408,groenendijk:2018p,wu:2019p,Kimbell:2020p054414,Ahadi:2017P172403,Ahadi:2018p056105} in ferromagnetic metal or from multiple Weyl nodes in itinerant antiferromagnets~\cite{Takahashi:2018p7880}. However, we rule out such possibilities in the present case, as oxygen-deficient KTO is a paramagnetic system. Furthermore, extracted ${R^\text{THE}_{xy}}$ of oxygen-deficient KTO shows all  characteristics features of THE, reported recently in paramagnetic phase of SrRuO$_3$ and V doped Sb$_2$Te$_3$~\cite{Wang:2019p1054}: ${R^\text{THE}_{xy}}$ is zero at $B$=0 and linear with $B$ at low field region ({\bf Figure}~\ref{fig:3}b-c). THE also decays as 1/$B^2$ in high field region (Figure S8, Supporting Information). Such dependency
on magnetic field can be successfully accounted by thermal fluctuations of scalar spin chirality in two dimension~\cite{Wang:2019p1054}. The microscopic origin of noncoplanar spin arrangement in our oxygen deficient KTO samples will be discussed later in the text. We also note that THE  is absent above 35 K, where WAL effect also vanishes. This observation implies that implies that the THE is contributed by only few ($\leq$ 16 nm ) nanometers layers from the top surface. This thickness is similar to the reported length scale for Rashba SOC-driven spin texture in 2D electron gas at the STO (111)
surface~\cite{He:2018p266802}. The magnitude of THE in our oxygen deficient KTO sample is also comparable with other systems from literature (see Table S1, Supporting Information). We have also checked our results by simulating  ${R^\text{AHE}_{xy}}$ with Langevin function which describes magnetization of a classical paramagnetic gas in an external field, and the conclusions remain the same (Figure S7, Supporting Information).

\begin{figure}
	\centering{
		\hspace*{-0.3cm}
		\includegraphics[scale=.7]{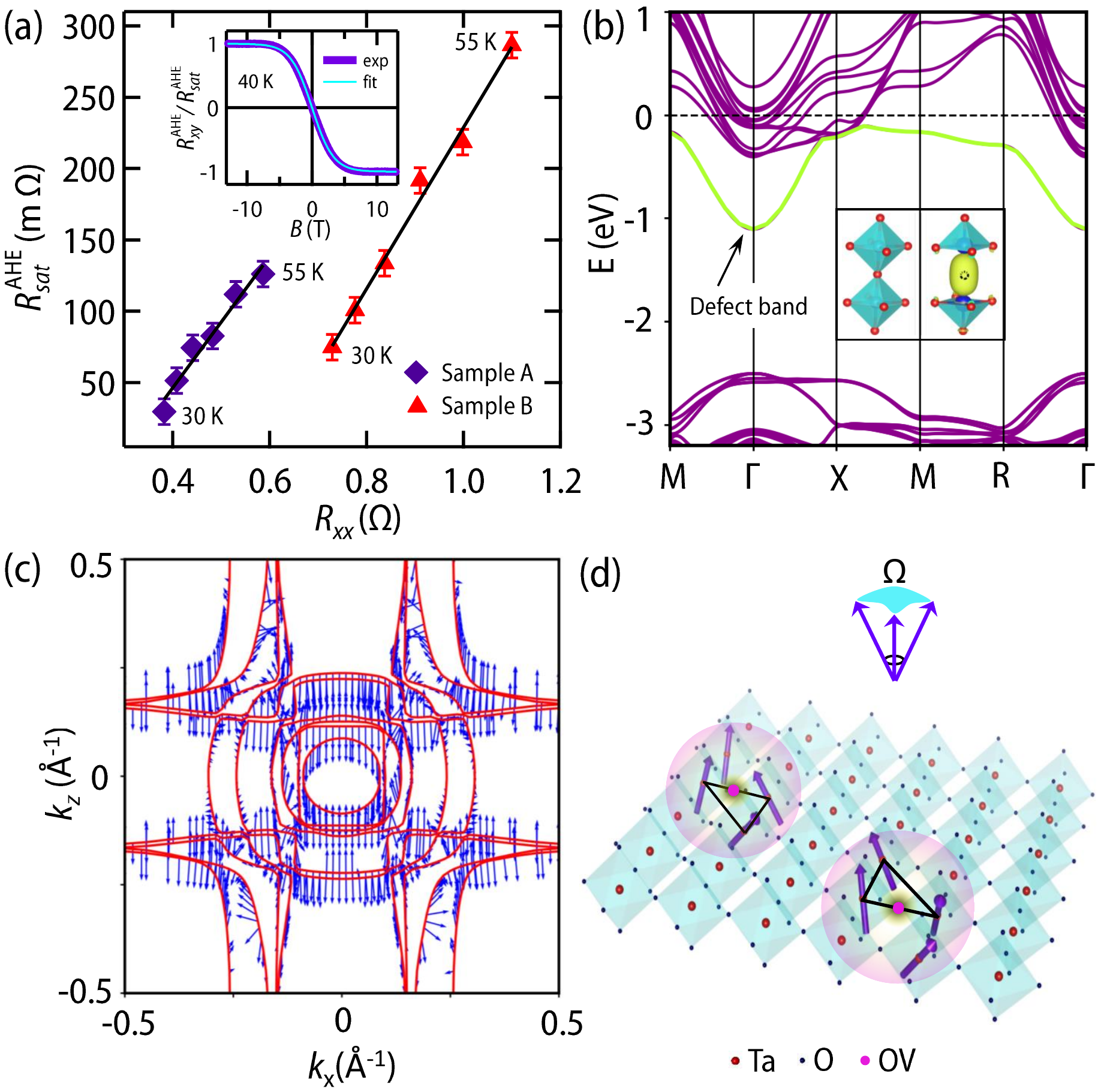}
		\caption{ Origin of THE and AHE. a) Scaling of ${R^\text{AHE}_{sat}}$ with $R$$_{xx}$ for both samples A and B. For this scaling analysis, we have focused in the temperature range where only AHE is present, error bar has been assigned to uncertainty in the ${R^\text{AHE}_{sat}}$ upon multiple measurements on the same sample. As evident, there is clear linear dependence which is a signature of skew scattering. Inset shows the fitting of normalized anomalous Hall resistance ${R^\text{AHE}_{xy}}$/${R^\text{AHE}_{sat}}$ with Brillouin function at 40 K. b)  Band structure of oxygen deficient KTO (KTaO$_{2.875}$) for a supercell of size $2 \cross 2 \cross 2$ with a single OV. The inset shows the isosurface plot of squared wave function of the defect band. c) Momentum space spin texture for $2 \cross 2 \cross 2$ supercell with single OV. The spins are projected into the $x$-$z$ plan. For the details about the calculations of these band resolved spin polarization vectors, we refer to the study by Altmeyer et al.~\cite{Altmeyer:2016p157203}. We have chosen vector scale of 10 to show this spin texture. d) Schematics of BMP with non collinear magnetic moments. Removal of oxygen leads to formation of localized moments on surrounding Ta atom. Moments inside a BMP become non collinear in presence of Rashba type SOC~\cite{denisov:2018p162409}. } \label{fig:4}}
\end{figure}

 The underlying  mechanism of AHE is very often discussed in terms of scaling of saturation value of ${R^\text{AHE}_{xy}}$ (${R^\text{AHE}_{sat}}$ ) with longitudinal resistance where ${R^\text{AHE}_{sat}}$ is proportional to some algebraic power of $R_{xx}$  (${R^\text{AHE}_{sat}}$ $\propto$ ${R_{xx}}^n$). Here $n$ can be integer or fraction and different values of $n$ corresponds to different scattering mechanism~\cite{nagaosa:2010p1539}.
 To examine this,  we plot the variation of ${R^\text{AHE}_{sat}}$ with $R$$_{xx}$ in {\bf Figure}~\ref{fig:4}a. As clearly evident, ${R^\text{AHE}_{sat}}$ scales linearly with $R$$_{xx}$, which establishes that AHE in our samples is dictated by skew scattering.
  Most importantly, we find that Brillouin function describes the field dependence of normalized AHE ${R^\text{AHE}_{xy}}$/${R^\text{AHE}_{sat}}$ for all temperatures very well (see inset of {\bf Figure}~\ref{fig:4}a and Figure S10a, Supporting Information). Local moment extracted from this fitting is found to be greater than 15 $\mu$$_B$ (see Figure S10b, Supporting Information). Similar value has also been obtained for the AHE of MgZnO/ZnO heterostructures~\cite{maryenko:2017p14777} and ultra-thin films of SrRuO$_3$~\cite{Wang:2019p1054}.

  The presence of local magnetic moments within our samples is essential for the manifestation of AHE and THE. To understand the origin of local moment in oxygen deficient  KTO, we have performed electronic structure calculations  by removing one O atom from a $2 \cross 2 \cross 2$ supercell. The total energy of the non-collinear configuration of the moments around the defect is lower by 54 meV/vacancy than a collinear one.
   The magnitude of the total magnetic moment is found to be 1.12  $\mu$$_B$/vacancy.  Out of the two electrons donated by the OV, one electron  is doped to the conduction band and the remaining electron results a completely occupied band ({\bf Figure}~\ref{fig:4}b), as also found for STO~\cite{Chungwei:2013p217601}.
   The charge density  of this fully occupied  band at $\Gamma$ point (inset of {\bf Figure}~\ref{fig:4}b) is highly localized around the OV, demonstrating that this band indeed arises from the defect. Not surprisingly, the defect band is composed mainly of Ta 5$d$ states from the Ta atom neighbouring the vacancy. This is further evident from the plot of total density of states (DOS) and the projected $d$ DOS of the Ta atoms situated away from the vacancy  and the Ta atoms adjacent to the vacancy (Figure S11, Supporting Information). In order to investigate the non-collinear nature of the spins near $E_F$, we have calculated the band resolved spin polarization vectors for the system containing OV. {\bf Figure}~\ref{fig:4}c shows the spin polarization vector for each of the bands crossing $E_F$ projected into the $x-z$ plane. The strong dependence of the spin polarization vector direction on ($k_x, k_z$) signifies the presence of Rashba-type SOC for the conduction electrons~\cite{Altmeyer:2016p157203}, further supporting our findings from the fitting of WAL.

      In order to explain the origin of THE and AHE in oxygen deficient KTO sample, we propose the following mechanism. Clustering of OVs is very likely, which would result localized moment on several Ta atoms around OVs. As well-known  in the context of dilute magnetic oxides,  local moments around a defect can form  bound magnetic polaron with large effective magnetic moment~\cite{coey:2005p173}. Moreover, the presence of conduction electrons with strong Rashba SOC (which is evident from WAL behavior and DFT calculations in present case) would further result non collinear BMP, as proposed recently by Denisov and Averkiev~\cite{denisov:2018p162409}. Such non collinearity within a BMP would give rise to non-zero scalar spin chirality, leading to the observation of THE ({\bf Figure}~\ref{fig:4}d). With the increase of $B$, all individual moments within a BMP will align along the direction of field, resulting zero THE for high $B$. As inferred from the WAL fitting, the effect of SOC becomes insignificant above 35 K, which destroys the noncollinearity within a BMP.  The asymmetric scattering of  conduction electrons from such collinear BMP can account for the observed skew scattering driven AHE above 35 K (Figure S12, Supporting Information) and the large moment obtained from the fitting of normalized AHE (inset of Figure 4a).

In summary, we have demonstrated emergent THE in  KTaO$_3$ through oxygen vacancy creation.  The observation of WAL and THE in similar temperature range  strongly implies definite role of Rashba type SOC behind the origin of THE. OVs induced magnetic moments act like asymmetric scattering centres in absence of Rashba type SOC, leading to the observation of sole AHE above 35 K.  Our present work demonstrates that defect engineering of a nonmagnetic, insulating system can result a metallic phase with  nontrivial spin textures, which can be further explored for topological spintronics~\cite{Fert:2017p17031}.

\section{Experimental Section}

 OVs in pristine KTO single crystal (dimension 5 mm$\times$5 mm$\times$0.5 mm)  have been incorporated by thermal annealing in presence of titanium wire within an evacuated sealed quartz tube by heating at $900 ^\circ$ C for 24 hours (see Supporting Information). Two samples were prepared with different vacuum level:  $\sim5 \times10^{-6}$ bar (sample A) and $10^{-3}$ (sample B).  All electrical transport measurements were carried out in four-probe Van der Pauw geometry using a Quantum Design physical property measurement system. Ohmic contacts were made using  ultrasonic bonding of aluminium wire as well as using gold wire with conductive silver paste and the results were same. For magnetoresistance (MR) and Hall measurements at fixed temperature, magnetic field  was varied between $\pm$13 T.  All MR vs. $B$ data have been plotted in main text after symmetrization with respect to $B$ 
 Similarly,  all $R_{xy}$ vs. $B$ data have been plotted after antisymmetrization.

\section{Supporting Information} 

Supporting Information is available.

\section{Acknowledgements} We are thankful to Prof. Mohit Randeria, Prof. Sumilan Banerjee and Prof. Ashis Kumar Nandy for fruitful discussions.  SKO and SM thank Professor D. D. Sarma for giving access to several experimental facilities for this work and Sayak Mondal for help with the sample preparation. This work is funded by a DST Nanomission grant (DST/NM/NS/2018/246), and a SERB Early Career Research Award (ECR/2018/001512). SM also acknowledges support from Infosys Foundation, Bangalore. The authors are grateful to Supercomputer Education and Research Centre, IISc for providing computational facilities.  The authors also acknowledge the experimental support from UGC-DAE CSR, Indore, in-mates of LT \& Cryogenics especially Er. P. Saravanan for their technical support and DST India for their initial support to 14 Tesla PPMS. V. G. contributed in this work  while in service.

\section{Conflict of Interest} The authors declare no conflict of interest.

\section{Keywords}  Topological Hall effect, anomalous Hall effect, spin-orbit coupling, magnetoresistance, weak antilocalization

\clearpage

	\renewcommand{\thefigure}{S\arabic{figure}}
	\renewcommand{\thetable}{S\arabic{table}}

	\makebox[\textwidth]{\bf \Large Supporting Information}
	
	\hspace{1cm}

	\makebox[\textwidth]{\bf \large Oxygen vacancy induced  topological  Hall effect in a nonmagnetic band insulator}
	
	\hspace{1cm}
	
	\makebox[\textwidth]{\large Shashank Kumar Ojha, Sanat Kumar Gogoi, Manju Mishra Patidar, Ranjan Kumar Patel,}
	\makebox[\textwidth]{\large  Prithwijit Mandal, Siddharth Kumar, R. Venkatesh, V. Ganesan, Manish Jain,  and Srimanta Middey}

	\hspace{1cm}
	
	{\bf \large A. \hspace{0.4 cm} Sample preparation and confirmation of oxygen vacancy:}
	
	To create oxygen vacancy (OV) in KTaO$_3$ (KTO), as received pristine KTO (001) substrate from Princeton Scientific Corp. was sealed in a quartz tube (Figure S1a) along with Ti wire under vacuum. This setup was then heated at 900 $^{\circ}$C for 24 hours. In this work, we are reporting data on two representative samples namely sample A and sample B. Same heating protocal was followed for both the samples, while only difference being the  vacuum during sealing. The vacuum during sealing for sample A was $\sim$ 5$\times$ 10$^{-6}$ bar and $\sim$ 10$^{-3}$ bar for sample B.
	
\setcounter{figure}{0}

	\begin{figure} [h]
		\vspace{-0pt}
		\includegraphics[width=0.75\textwidth] {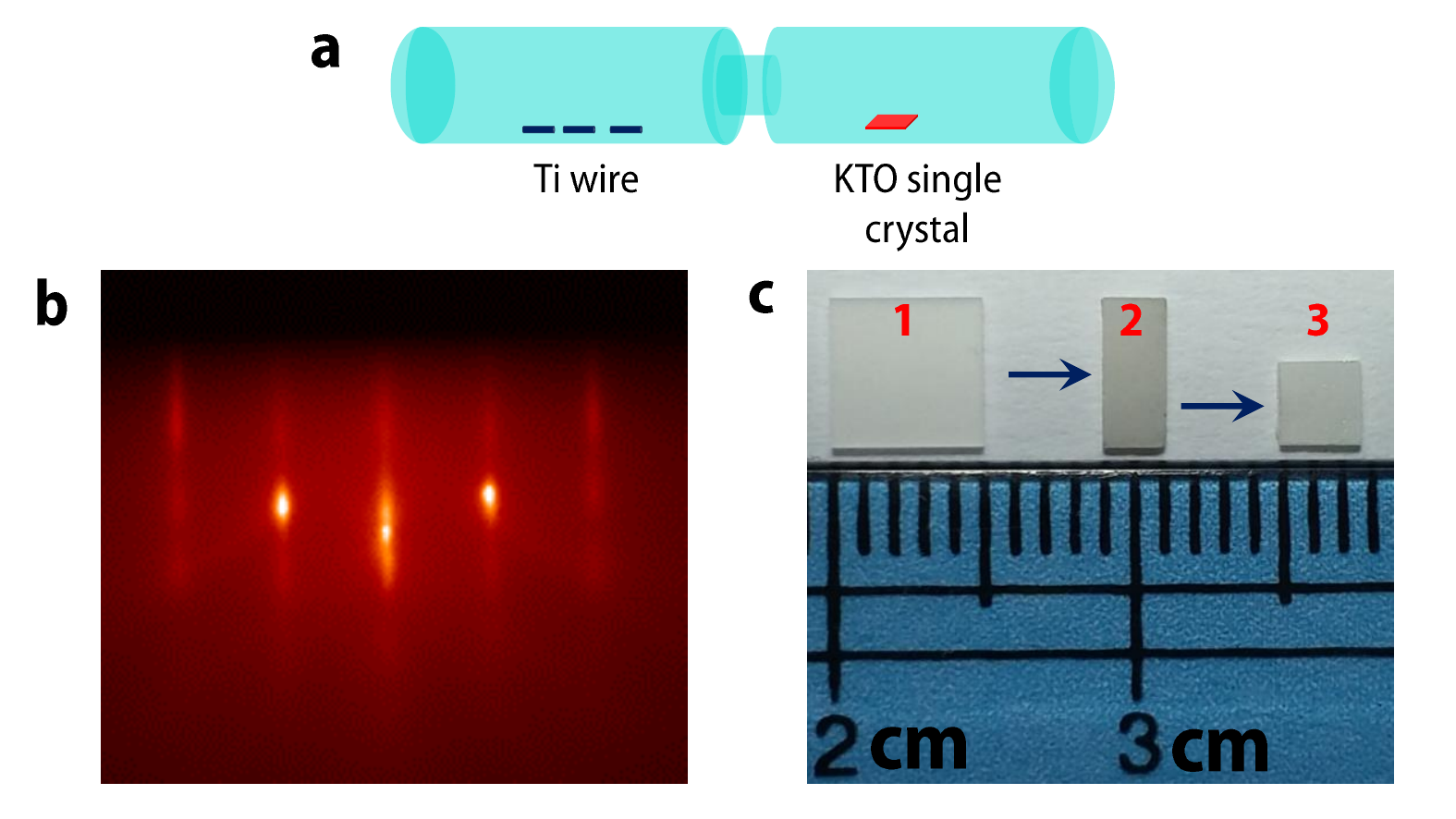}
		\caption{\label{FigS1} \textbf{a.} Quartz tube sealing arrangement (Ti wire has been shown in the left side  and KTO single crystal has been shown in the right). \textbf{b.} RHEED pattern of oxygen deficient KTO. \textbf{c.} Crystal marked by number 1 in red color is the image of as received transparent substrate, sample marked by number 2 is the image after oxygen vacancy creation (crystal turns slight slaty gray). Upon heating oxygen deficient KTO in high pure oxygen, it becomes transparent similar to pristine KTO (see crystal marked by number 3). }
	\end{figure}
	
	This method of creating OV is superior than argon ion bombardment (AIB) as AIB destroys the sample and makes it amorphous~\cite{kan:2005p816}. Crystallinity of our sample was confirmed by Reflection High Energy Electron Diffraction (RHEED). Figure S1b shows the RHEED image of oxygen deficient KTO, intense specular and off specular spots confirm that the KTO still remains crystalline even after OVs.

	To testify that the metallic beahavior is related to OV, we have also heated one oxygen deficient KTO sample in high pure oxygen for 24 hours. The oxygen deficient KTO sample becomes transparent similar to pristine KTO after oxidation (Figure S1c). Also, the resistance of oxygen deficient KTO becomes immeasurable after oxidation, which confirms oxygen vacancy induced metallicity in our samples.

	\clearpage

	{\bf \large B. \hspace{0.4 cm} Extraction of effective mass from SdH oscillations:}
	
	\begin{figure} [h]
		\vspace{-0pt}
		\includegraphics[width=0.7\textwidth] {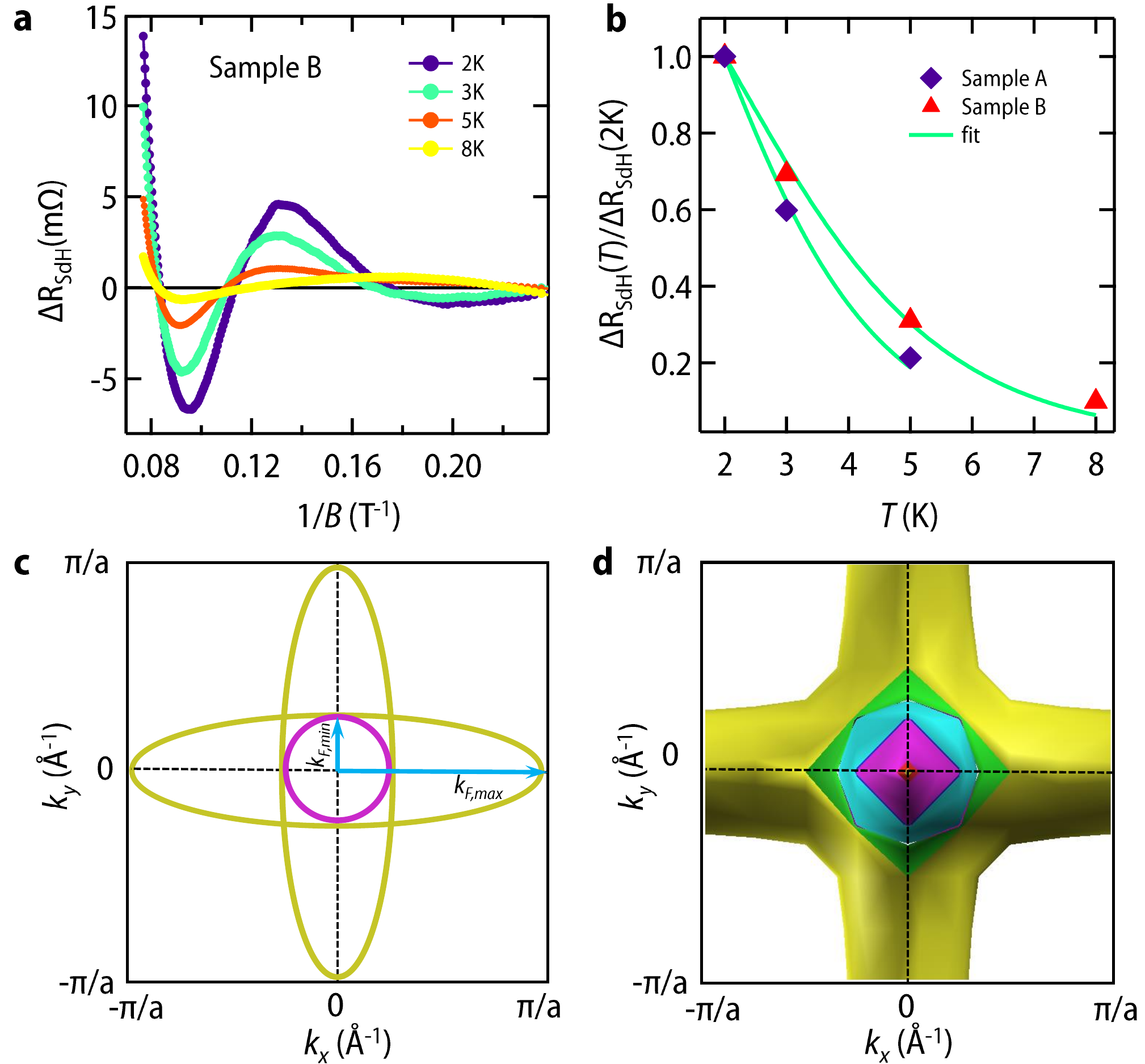}
		\caption{\label{FigS2} \textbf{a.} Extracted SdH amplitude for sample B obtained by subtracting 2$^\text{nd}$ order polynomial from $R_{xx}$. \textbf{b.} Fitting of temperature dependent scaled oscillation amplitude for extraction of effective mass. \textbf{c.} Idealized Fermi surface of electron doped KTO. \textbf{d.} Calculated Fermi surface for KTO with single oxygen vacancy in 2$\times$2$\times$2 supercell.}
	\end{figure}
	
	The effective mass of electron can be extracted by fitting temperature dependent SdH amplitude with the following formula~\cite{shoenberg:2009magnetic}
	\begin{align}
	\dfrac{\Delta \text{R}_{\text{SdH}}(T)}{\Delta \text{R}_{\text{SdH}}(T_0)} = \dfrac{T \text{sinh}(\alpha T_0)}{T_0 \text{sinh}(\alpha T)}  \label{eq:1}
	\end{align}
	where $\alpha$ = 2{${{\pi^2}k_B/{\hbar \omega _c}}$}, $\omega _c$ = $e$$B$/${m^{\star}}$ is the cyclotron frequency, $m^{\star}$ is the effective mass, $\hbar$ is the reduced Planck's constant and $T_{0}$ is 2 K. Fitting leads to an effective mass of $m^{\star}$=(0.56$\pm$0.02)$m_{e}$ for sample A and  $m^{\star}$=(0.48$\pm$0.02)$m_{e}$ for sample B ( here $m_{e}$ is the bare electron mass) (see Figure S2b). The value of effective mass is in good agreement with previous literature~\cite{harashima:2013p085102}. Furthermore, this value of $m^{\star}$ yields  the Landau level (LL) energy splitting of $\hbar\omega$ $_c$ = $\hbar$ $e$$B$/${m^{\star}}$ $\approx 0.85 meV$ for sample A. This energy is equivalent to $\sim$ 10 K, explaining the absence of oscillations above 10 K due to the thermal broadening of LL.

	We now proceed to determine the thickness of the electron gas which would involve several assumptions as discussed below. For this, we first briefly discuss here Fermi surface topology of electron doped KTO. The Fermi surface of KTO~\cite{mattheiss:1972p4718} is very similar to SrTiO$_3$ and comprises of three ellipsoids of revolution centered at $\Gamma$ point with major axis $k_{F,max}$ along [100] crystallographic axis and the minor axis ${k_{F,min}}$ transverse to it. In the presence of magnetic field along [001] crystallographic axis, electrons traverse around the extremal orbits as shown in the Figure S2c. The extremal orbits corresponding to the two ellipsoids directed along x and y axis are ellipses with  area equal to $\pi$ $k_{F,min}$ ${k_{F,max}}$, whereas the extremal orbit corresponding to the ellipsoid along the z direction is circular with an area ${\pi k^{2}_{F,min}}$. Also from literature it is known that $k_{F,max}$/$k_{F,min}$=1.541~\cite{uwe:1979p3041,mattheiss:1972p4718}, which implies that there are more electrons around elliptical orbit than those orbiting around circular orbit. From above arguments, we assume that main frequency peak of SdH oscillation comes from electrons orbiting around the elliptical orbit. Also it is known that, frequency of SdH oscillation F$_\text{SdH}$ is given by F$_\text{SdH}$=$\hbar$/{2A$_{ext}$$e$}~\cite{herranz:2007p216803} where A$_{ext}$ is the area of extremal orbit in the $k$ space and -$e$ is the electrons charge. Since  F$_\text{SdH}$ is known (12.8 T for sample A and 11 T for sample B), one can calculate product of $k_{F,max}$ and $k_{F,min}$ from here. From above analysis one can now calculate $k_{F,max}$ and $k_{F,min}$ individually and carrier density can be determined. For this, we note that the total volume occupied by three ellipsoid is V${_k=3\times(4\pi/3){k^2}_{F,min}k_{F,max}}$ from which the carrier concentration $n_v$ can be calculated as $n_v$=${({k^2}_{F,min}k_{F,max})/{\pi^2}}$. Now defining the sheet carrier density $n$$_{s}$ as product of carrier density $n_v$ and the thickness of conducting region $t$ one can write $n_s$=$n_v\times t$. From Hall measurement ${n_s}$ is known (see section \textbf{E}) which results in a thickness of about 35 microns and 43 microns for sample A and B respectively. Since this thickness is much larger than Fermi wavelength, our system is 3 dimensional in nature. Vary similar analysis for STO determines the metallic region to be several hundred microns~\cite{herranz:2007p216803}.
	
	\clearpage
	
	{\bf \large C. \hspace{0.4 cm}  Magnetoresistance (MR) from 2 K to 300 K:}
	
	Figure S3 shows the systematic evolution of MR from 2 K to 300 K for the sample A and B. As discussed in the main  text and section B of this Supporting Information, MR shows SdH oscillations below 10 K for $B$ $>$ 5 Tesla. Similar to sample A, weak antilocalization is also present in sample B which vanishes above $\sim$ 50 K. Above 50 K, MR is almost linear at high field and has also small $B$$^2$ contribution.
	\begin{figure} [h]
		\vspace{-0pt}
		\includegraphics[width=1.0\textwidth] {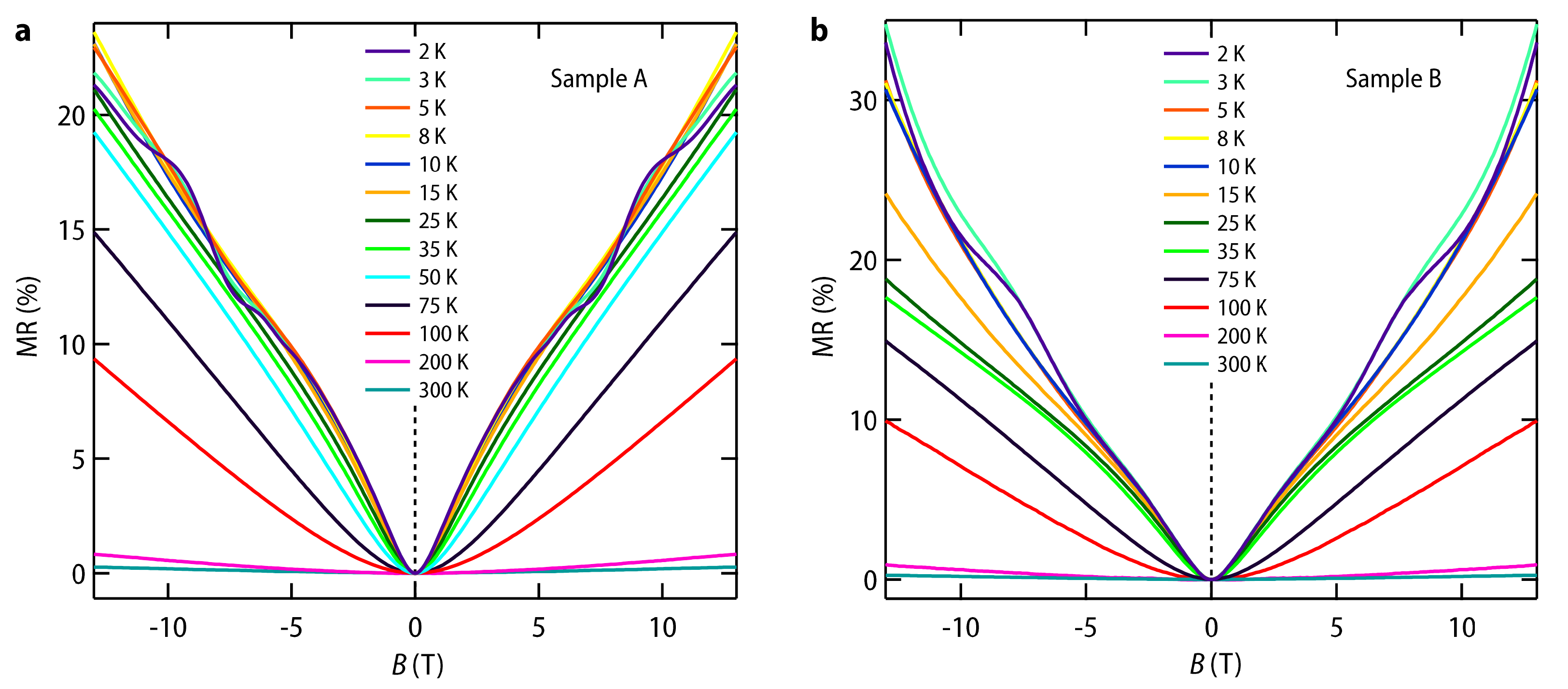}
		\caption{\label{FigS} \textbf{a.} MR for sample A. \textbf{b.} MR for sample B. }
	\end{figure}
	
	\clearpage
	
	{\bf \large D. \hspace{0.4 cm} Two band model fitting of $R_{xy}$:}
	\begin{figure} [h]
		\vspace{-0pt}
		\includegraphics[width=1\textwidth] {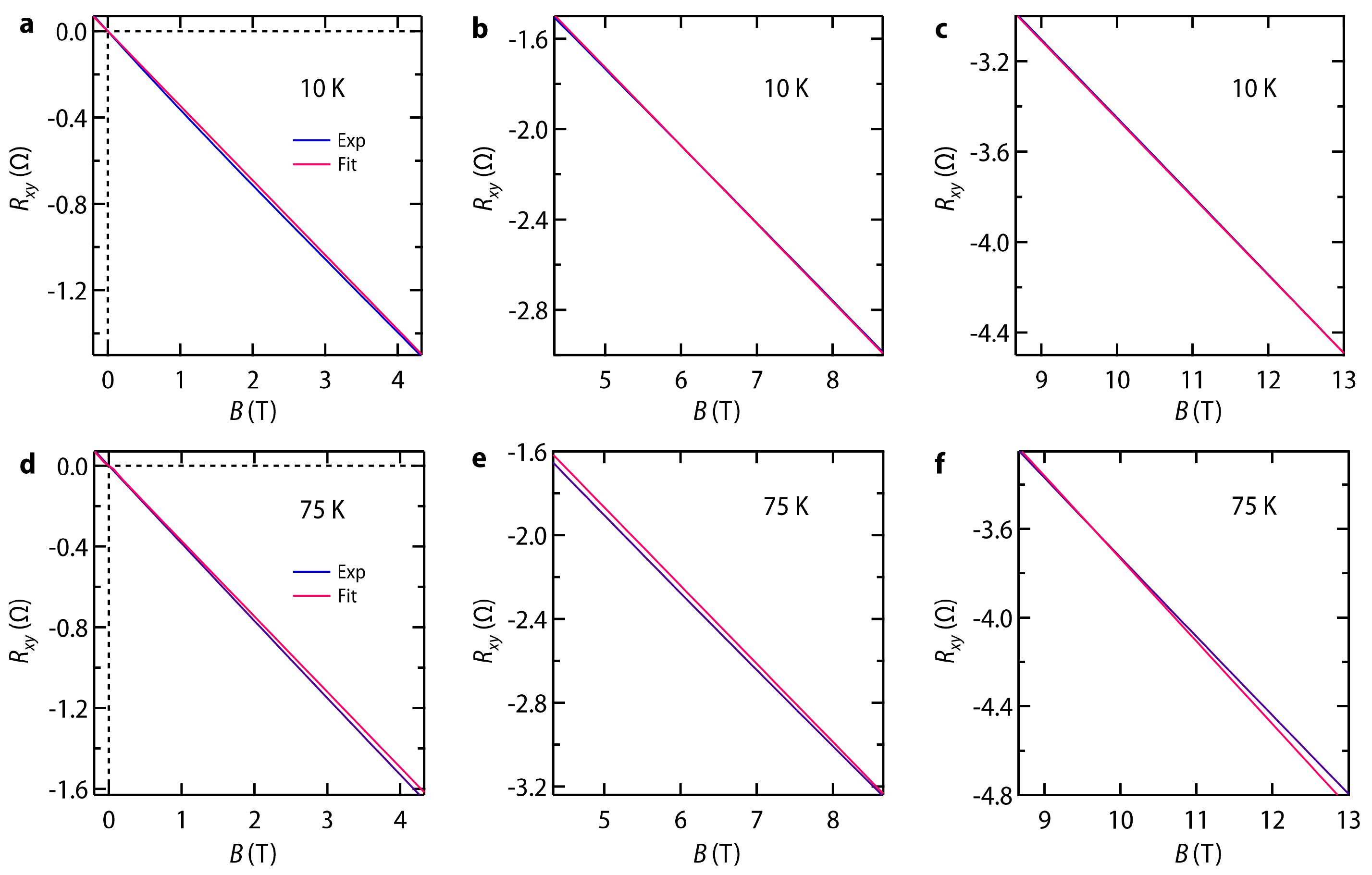}
		\caption{\label{FigS4} For visual clarity, whole $R_{xy}$ along with fitting with two band model  has been plotted in three panels from 0 T to 13 T. \textbf{a. b. c.} Fitting of $R_{xy}$ with two band model at 10 K. \textbf{d. e. f.}  Fitting of $R_{xy}$ with two band model at 75 K.}
	\end{figure}
	
	Non linear Hall effect has been very often associated with two band transport. To check for this, antisymmetrized $R_{xy}$ was fitted with two band model~\cite{gunkel:2016p031035}
	
	\begin{align}
	R_{xy}= -\frac{1}{e}{\dfrac{(\frac{n_1\mu_1^2}{1+\mu_1^2B^2}+\frac{n_2\mu_2^2}{1+\mu_2^2B^2})B}{(\frac{n_1\mu_1}{1+\mu_1^2B^2}+\frac{n_2\mu_2}{1+\mu_2^2B^2})^2+(\frac{n_1\mu_1^2}{1+\mu_1^2B^2}+\frac{n_2\mu_2^2}{1+\mu_2^2B^2})^2B^2}}  \label{eq:3}
	\end{align}

	assuming electrons in both the bands, with the constraint
	\begin{align}
	R_{S}^{-1}(B=0)= e(n_1\mu_1+n_2\mu_2).
	\end{align}
	Here $e$ is the magnitude of charge of electron, $n_1$ \& $n_2$ are the sheet carrier density corresponding to band 1 and band 2 respectively, $\mu_1$ \& $\mu_2$ are the mobility corresponding to band 1 and band 2 respectively and $R_S$ is the sheet resistance. Figure S4 shows fitting of antisymmetrized $R_{xy}$ at two representative temperatures  10 K and 75 K for sample A. We would like to emphasize that two band model does not fit our data in the field range where there is non linearity in $R_{xy}$. For example, at 10 K, two band model does not fit upto $\sim$ 4 T (at 10 K, $R_{xy}$  is non-linear upto ~ 5 T and becomes linear at high field as evident from $\frac{dR_{xy}}{dB}$ (see Figure 2b of main text)) and at 75 K it does not fit in the whole field range from 0 T to 13 T ( at 75 K non-linearity is present in whole field range as evident from $\frac{dR_{xy}}{dB}$ (see Figure 2b of main text)). This strongly suggests that, two band model is unable to capture non-linear Hall effect in present case.
	
	\clearpage
	
	{\bf \large E. \hspace{0.4 cm} Extraction of sheet carrier density and mobility:}
	
	Figure S5a shows the derivative of antisymmetrized $R_{xy}$ for sample A. As clearly evident, from 2 K to 50 K, derivative becomes field independent above 10 T and all the data merge together at highest field.  This observation demonstrates that the transport is effectively one band  as the $\frac{dR_{xy}}{dB}$ would be a function of $B$ for two band model. Similar trend has been also observed for the sample B (Figure S5b). One band model results a temperature independent (upto 50 K) sheet carrier density $n_s$ $\sim$ 1.8 $\times$ 10$^{15}$ cm$^{-2}$ for sample A and ~$\sim$ 1.0 $\times$ 10$^{15}$ cm$^{-2}$ for sample B. Low temperature mobility is around 2400 cm$^2$/V-s for both the samples. $n_s$ was calculated by the formula $n_s$=-1/$e$($\frac{dR_{xy}}{dB}$) and the mobilty was calculated by the formula 1/$R_s$ = $n_s$$e$$\mu$. We could not extract $n_s$ and $\mu$ above 50 K due to the fact that $\frac{dR_{xy}}{dB}$ is strongly field dependent upto 13 Tesla.

	\begin{figure} [h]
		\vspace{-0pt}
		\includegraphics[width=0.9\textwidth] {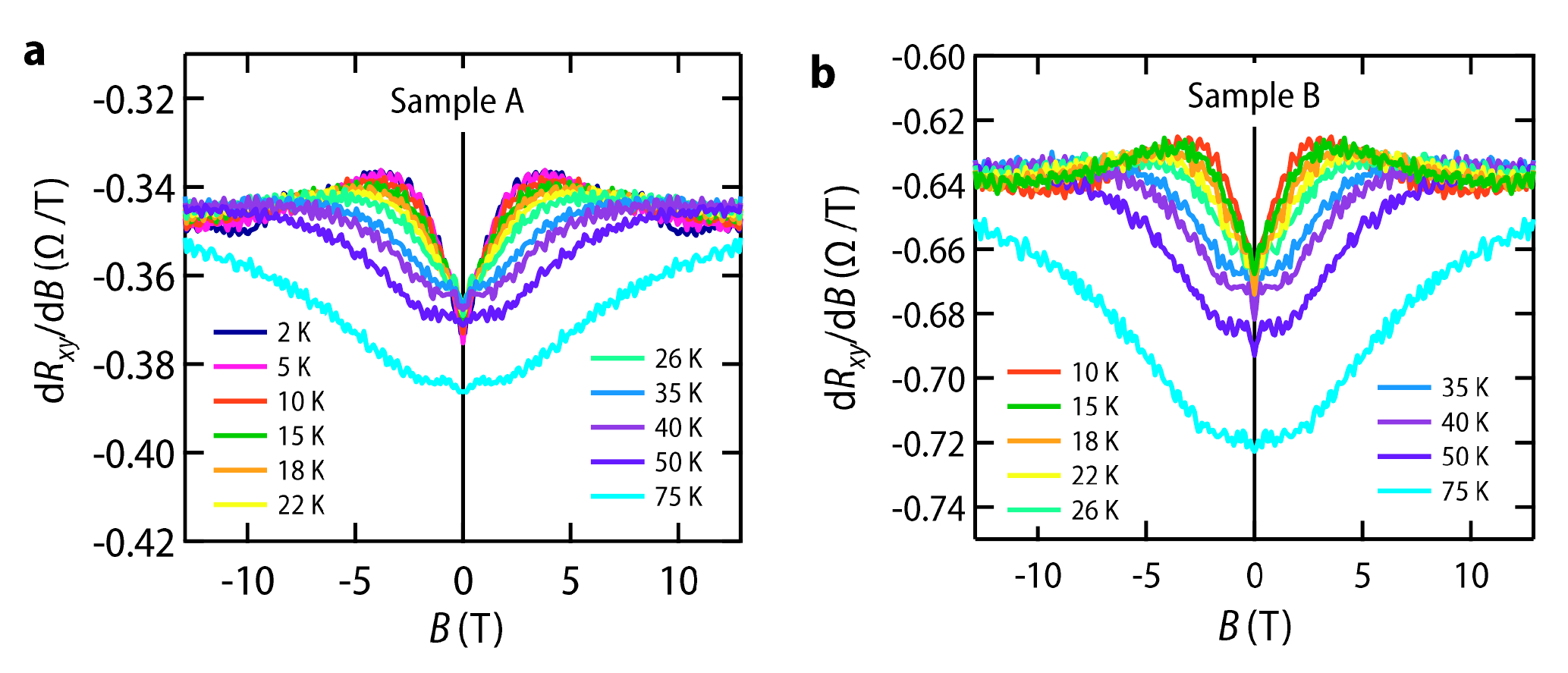}
		\caption{\label{FigS5} \textbf{a.} Derivative of antisymmetrized $R_{xy}$ from 2 K to 75 K for sample A. \textbf{b.} Derivative of antisymmetrized $R_{xy}$ from 10 K to 75 K for sample B. }
	\end{figure}
	
	\clearpage
	
	{\bf \large F. \hspace{0.4 cm} Non-linearity in Hall resistance $R$$_{xy}$ of sample B:}
	
	Figure S6a shows the  anti-symmetrized Hall resistance $R$$_{xy}$ from 2 K to 300 K. The non-linearity (similar to sample A) in $R$$_{xy}$ in the form of dip is evident from the $\frac{dR_{xy}}{dB}$ shown in Figure S6b. With increase in temperature this non linearity spreads at higher fields and eventually derivative becomes totally field independent at 300 K. Interestingly, $\frac{dR_{xy}}{dB}$ at 2 K and 5 K even show oscillations
	at high field. Similar oscillation is Bi$_2$Se$_3$ and Bi$_2$Te$_3$
	was attributed to the onset of quantum Hall regime~\cite{busch:2018p485,qu2010:p821}.
	
	FIG. S6c shows the ${R^\text{extra}_{xy}}$ (${R^\text{AHE}_{xy}}$+${R^\text{THE}_{xy}}$) from 10 K to 55 K. High temperature ${R^\text{extra}_{xy}}$ is characterized by pure AHE similar to sample A. Hump like feature below 40 K clearly shows the presence of THE. Temperature evolution of THE component for sample B has been plotted in Figure S9.
	
	\begin{figure} [h]
		\vspace{-0pt}
		\includegraphics[width=0.75\textwidth] {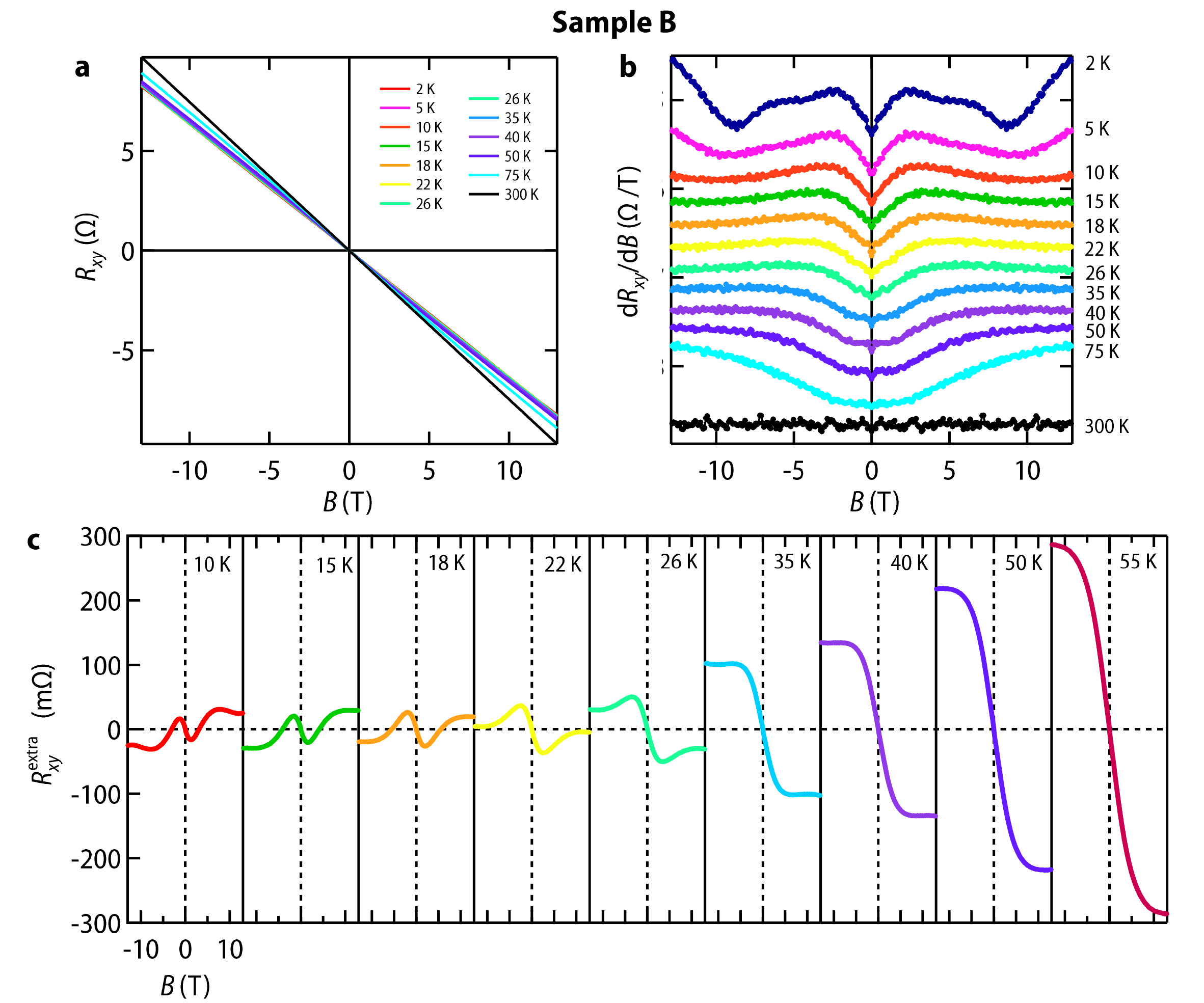}
		\caption{\textbf{a.}  Magnetic field dependent anti-symmetrized Hall resistance $R$$_{xy}$ measured at different temperatures for sample B. \textbf{b.} Derivative of $R$$_{xy}$ w.r.t applied magnetic field, all data have been shifted vertically for visual clarity. \textbf{c.} ${R^\text{extra}_{xy}}$ as a function of magnetic field $B$  at various temperatures from 10 K to 55 K. We could not do OHE subtraction above 55 K as $\frac{dR_{xy}}{dB}$ does not become field independent even at high fields. Also we could not do OHE subtraction below 10 K due to oscillations in $\frac{dR_{xy}}{dB}$ below 10 K .}
	\end{figure}

	\clearpage
	
	{\bf \large G. \hspace{0.4 cm} Extraction of ${R^\text{AHE}_{xy}}$ from ${R^\text{extra}_{xy}}$ :}

	As AHE is proportional to the magnetization of the sample, we have used Brillouin function to simulate ${R^\text{AHE}_{xy}}$. Brilluoin function describes the shape of magnetization of quantum paramagnetic gas in an external magnetic field  and is given by

	\begin{align}
	B_J(x)=\frac{2J+1}{2J}\text{coth}(\frac{2J+1}{2J})-\frac{1}{2J}\text{coth}(\frac{1}{2J}x)
	\end{align}
	\begin{align}
	x=\frac{g\mu_BJB}{k_BT}
	\end{align}

	\begin{figure} [h]
		\vspace{-0pt}
		\includegraphics[width=0.9\textwidth] {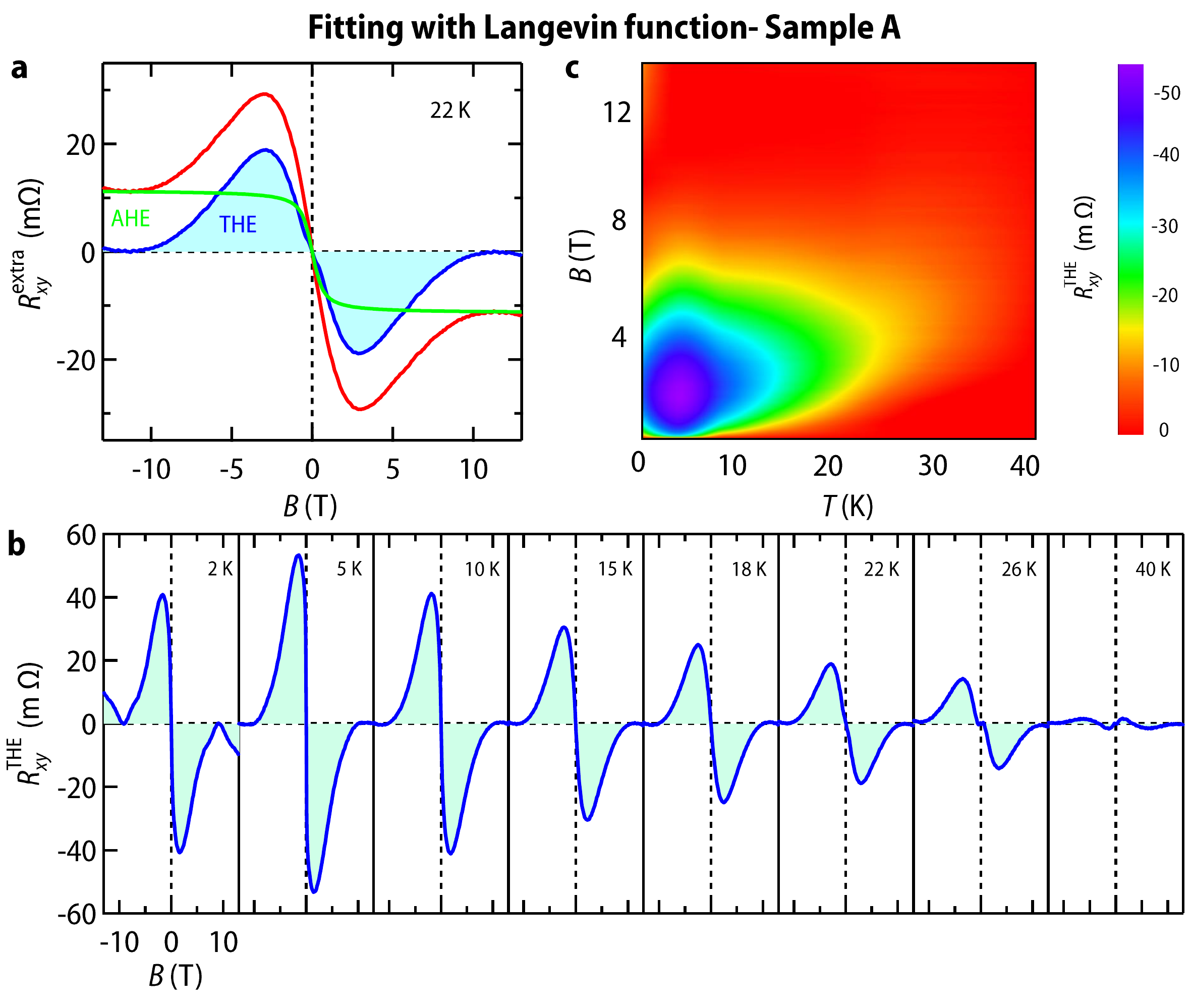}
		\caption{\label{FigS7}  \textbf{a.} Red curve shows the ${R^\text{extra}_{xy}}$ at 22 K, green curve is the fitting with a Langevin for estimation of  ${R^\text{AHE}_{xy}}$, blue curve filled with light sky color shows extracted ${R^\text{THE}_{xy}}$ obtained by subtracting ${R^\text{AHE}_{xy}}$ from ${R^\text{extra}_{xy}}$. \textbf{b.} Extracted ${R^\text{THE}_{xy}}$ as a function of magnetic field $B$ at various temperature from 2 K to 40 K. There is anomaly in ${R^\text{THE}_{xy}}$ at high field at 2 K due to   oscillations in ${R^\text{extra}_{xy}}$ (see Figure 2b of main text). \textbf{c.} Color map of ${R^\text{THE}_{xy}}$ in the $B$-$T$ plane. }
	\end{figure}

	where $J$ is total angular momentum quantum number, $g$ is Lange-$g$ factor, $\mu_B$ is Bohr magneton, $k_B$ is Boltzman constant, $T$ is temperature and $B$ is applied magnetic field. To simulate ${R^\text{AHE}_{xy}}$, ${R^\text{extra}_{xy}}$ was first normalized with saturation value of ${R^\text{extra}_{xy}}$ and then ${R^\text{AHE}_{xy}}$ was simulated (see Figure 3a of main text.). We have assumed $g$=2 and $J$=1/2 (we have used these values of $J$ and $g$ because Brillouin function provides an excellent fit to our normalized  AHE assuming $g$=2 and $J$=1/2 (see section \textbf{J})). Further, $\mu_B$ was replaced with $\mu_{eff}$ and was kept as fitting parameter. Similar method was used to fit AHE for ZnO/MgO~\cite{Maryenko2017}. We have also used Langevin function to estimate ${R^\text{AHE}_{xy}}$ and the conclusions remian the same (see Figure S7).
	
	\clearpage
	
	{\bf \large H. \hspace{0.4 cm}
		High field beahvior of ${R^\text{THE}_{xy}}$ :}
	Figure S8a-b shows the  ${R^\text{THE}_{xy}}$ at 10 K and 22 K for sample A. As clearly seen, high field THE signature decays as $\frac{1}{B^2}$. Similar trend has been observed in SrRuO$_3$ films and was attributed to fluctuations of the scalar spin chirality in two dimension~\cite{Wang:2019p1054}.
	\begin{figure} [h]
		\vspace{-0pt}
		\includegraphics[width=0.9\textwidth] {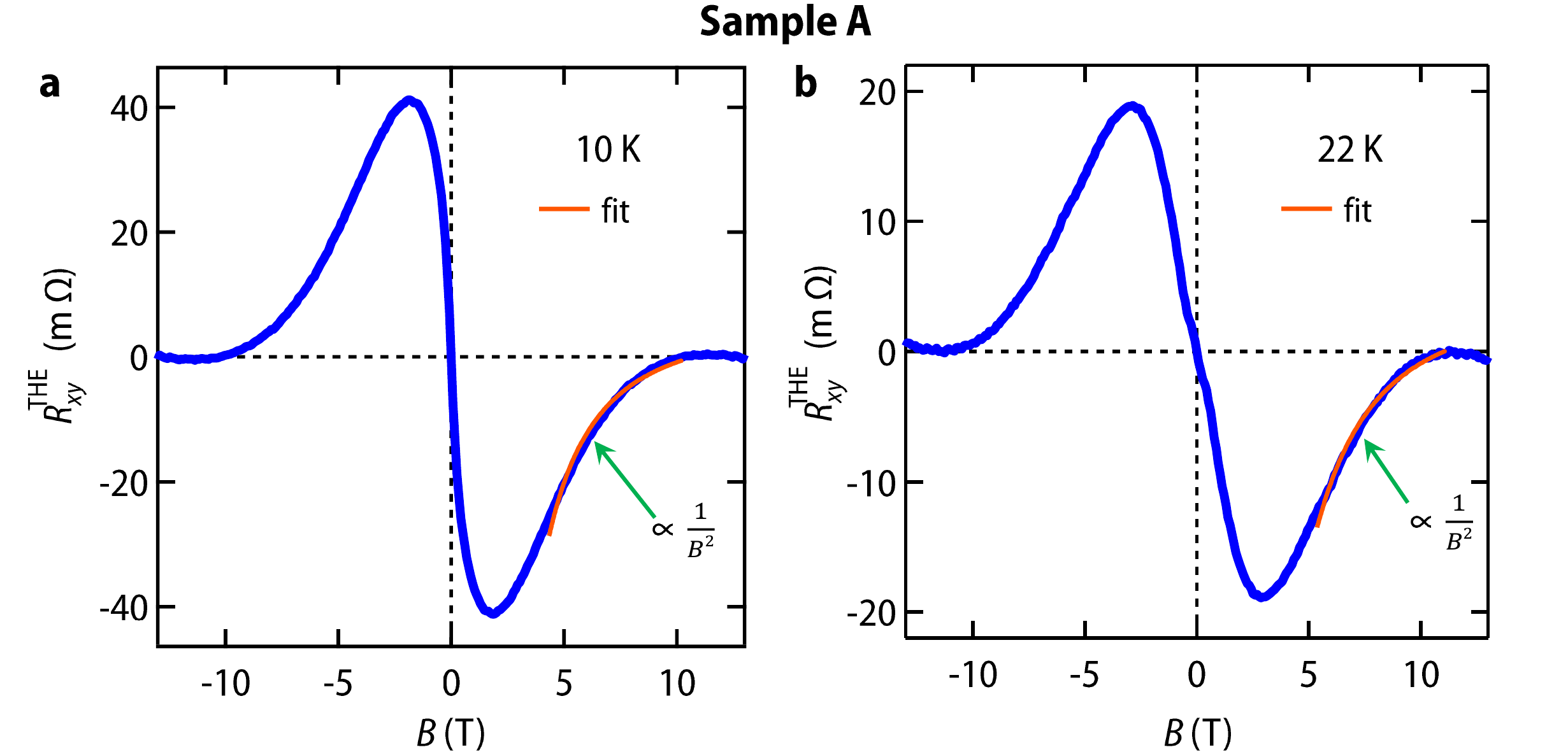}
		\caption{\label{FigS8}  \textbf{a.}  ${R^\text{THE}_{xy}}$ at 10 K along with high field fitting with a+b/$B^2$.\textbf{b.} ${R^\text{THE}_{xy}}$ at 10 K along with high field fitting with a+b/$B^2$.}
	\end{figure}

	\clearpage
	
	{\bf \large I. \hspace{0.4 cm} Temperature evolution of THE for sample B:}
	
	Figure S9 shows the extracted ${R^\text{THE}_{xy}}$ from 10 K to 40 K for sample B. Brillouin function was used for the estimation of ${R^\text{AHE}_{xy}}$ in this case. As evident there is clear signature of THE below 40 K.  We could not extract ${R^\text{THE}_{xy}}$ below 10 K due to the fact that ${R^\text{extra}_{xy}}$ itself can not be extracted below 10 K due to presence of oscillations in ${R^\text{extra}_{xy}}$ (see Figure S6b).
	
	\begin{figure} [h]
		\vspace{-0pt}
		\includegraphics[width=1\textwidth] {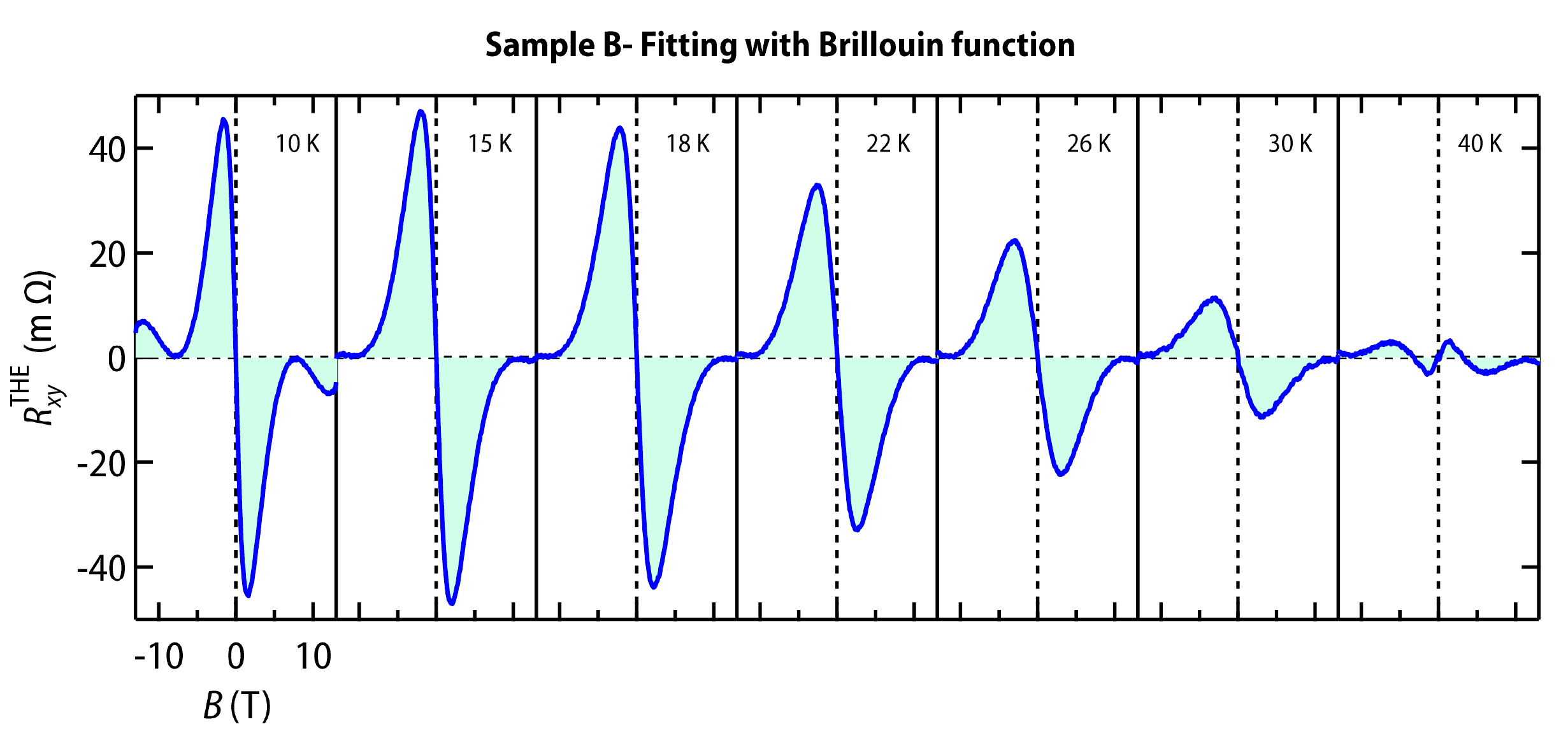}
		\caption{\label{FigS9} Extracted ${R^\text{THE}_{xy}}$ as a function of magnetic field $B$ at various temperature from 10 K to 40 K.}
	\end{figure}
	
	\clearpage
	
	{\bf \large J. \hspace{0.4 cm} Comparison of maximum amplitude of THE:}
	
	In the Table S1 we have compared the maximum amplitude of topological Hall effect for our sample with other systems from literature. For calculation of maximum value of THE for our samples we have used the thickness of 16 nm which is the maximum thickness of the surface layers contributing to the emergence of the THE.
	
	\begin{table}[h]
		\begin{center}
			\begin{tabular}{c|c|c}
				\hline
				\text{Material system} & \text{Maximum THE} & \text{Reference} \\
				&  ($\mu$$\Omega$ cm) & \\
				\hline
				MnSi  &   0.0045  &  {\it Phys. Rev. Lett.} {\bf 102}, 186602 (2009)\\
				MnP  & 0.01  &  {\it Phys. Rev. B} {\bf 86}, 180404 (2012)\\
				SrRuO$_3$/SrTiO$_3$  & 0.07  &  {\it Nat. Mater.} {\bf 18}, 1054 (2019)\\
				{\bf KTaO$_{\bf{3-\delta}}$}  &   {\bf 0.10}  & {\bf This work}\\
				MnGe & 0.16 &  {\it Phys. Rev. Lett.} {\bf 106}, 156603 (2011)\\
				SrRuO$_3$/SrIrO$_3$ & 0.2 & {\it Sci. Adv.} {\bf 2}, e1600304 (2016) \\
				Single layer SrRuO$_3$ & 0.2 & {\it Adv. Mater.} {\bf 31} 1807008 (2019) \\
				La$_{0.7}$Sr$_{0.3}$Mn$_{0.95}$Ru$_{0.05}$O$_3$ & 0.6 & {\it J. Phys. Soc. Jpn.} {\bf 87}, 074704 (2018) \\
				Eu$_{1-x}$La$_x$TiO$_3$  & 6 & {\it Sci. Adv.} {\bf 4}, eaar7880 (2018) \\
				\hline
			\end{tabular}      
			\caption{Comparison of maximum amplitude of THE. }
			\label{tab:table1}
		\end{center}
	\end{table}
	
	\clearpage
	
	{\bf \large K. \hspace{0.4 cm} Fitting of ${R^\text{AHE}_{xy}}$ with Brillouin function:}
	
	As discussed earlier, the Brillouin function provides an excellent fit to our normalized AHE  ${R^\text{AHE}_{xy}}$/${R^\text{AHE}_{sat}}$ at all temperatures  assuming $g$=2 and $J$=1/2 (this assumption works very well for systems with defect induced local moment~\cite{Maryenko2017}).  In Figure S10a we show the fitting of normalized AHE at four representative temperatures for sample A.  In Figure S10b  we show the extracted moment from fitting for both the samples. As evident, extracted moment is greater than 15 $\mu_B$.
	
	\begin{figure} [h]
		\vspace{-0pt}
		\includegraphics[width=1\textwidth] {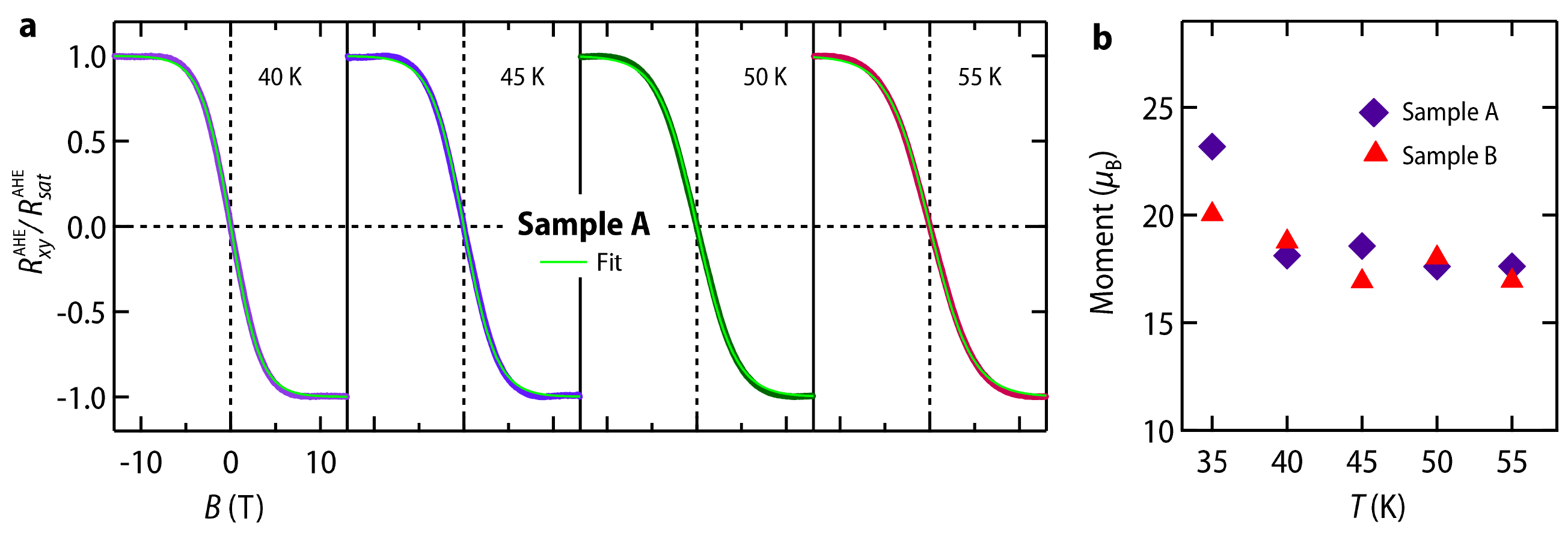}
		\caption{\label{FigS10} \textbf{a.}  Fitting of normalized Hall resistance ${R^\text{AHE}_{xy}}$/${R^\text{AHE}_{sat}}$ with Brillouin function for sample A. \textbf{b.} Extracted moment from the fitting for both the samples. }
	\end{figure}
	
	\clearpage
	
	{\bf \large L. \hspace{0.4 cm} Theoretical calculations:}
	
	The first-principle density functional theory (DFT) calculations for KTaO$_3$ and KTaO$_{3-\delta}$ have been carried out using the QUANTUM ESPRESSO package~\cite{QE-2017}.  We have employed Perdew, Burke, Ernzerhof  generalized gradient approximation~\cite{PBE-GGA} for the exchange corelation functional. The electron-ion interactions are described via norm-conserving pseudopotentials~\cite{oncv:scalar}. In order to capture the effect of SOC we have used fully relativistic pseudo-potentials for  Ta atoms.  Throughout the calculations, the electronic wave functions are expanded in plane waves with energy upto 90 Ry.
	For the unit cell calculations of pristine KTO, an optimized $8 \cross 8 \cross 8$ $k$-point grid sampling of the Brillouin zone is used. 
	The optimized lattice parameter is found to be 4.021 \si{\angstrom}~\cite{Choi:2011p214107}. The relaxed lattice parameter is in good agreement with the experimental value (3.989 \si{\angstrom})~\cite{wemple:1965pA1575} . We have checked that our conclusions do not change upon inclusion of an onsite Coulomb potential qualitatively. The pristine KTO is insulating in nature and it has a indirect band gap between $\Gamma$ and $R$ point. In our calculations we have found the indirect band gap to be 2.05 eV, which is in good agreement  with previous DFT results~\cite{kto:strain,kto:ktl,shing:kto}.
	The defect calculations are performed on supercells of sizes $ 2\cross  2\cross 2$ and $3 \cross 3 \cross 3$ for which the Brillouin zone is sampled with  $4 \cross 4 \cross 4$ and $2 \cross 2 \cross 2$ $k$-points, respectively.  We have checked that upon increasing the $k$-grid sampling  for the $2 \cross 2 \cross 2$ supercell the conclusions remain unchanged. The atoms in the supercell are relaxed until the forces are less than 0.07 eV/\si{\angstrom}.
	
	\clearpage
	
	{\bf \large M. \hspace{0.4 cm} DOS plot of oxygen deficient KTO:}
	
	Figure S11 shows the DOS plot of oxygen deficient KTO for a supercell of size 2$\times$2$\times$2 with a single OV. The defect states have been marked by black arrow. As evident, defect states are composed mainly of Ta 5$d$ states from the Ta atom neighbouring the oxygen vacancy.
	
	\begin{figure} [h]
		\vspace{-0pt}
		\includegraphics[width=0.7\textwidth] {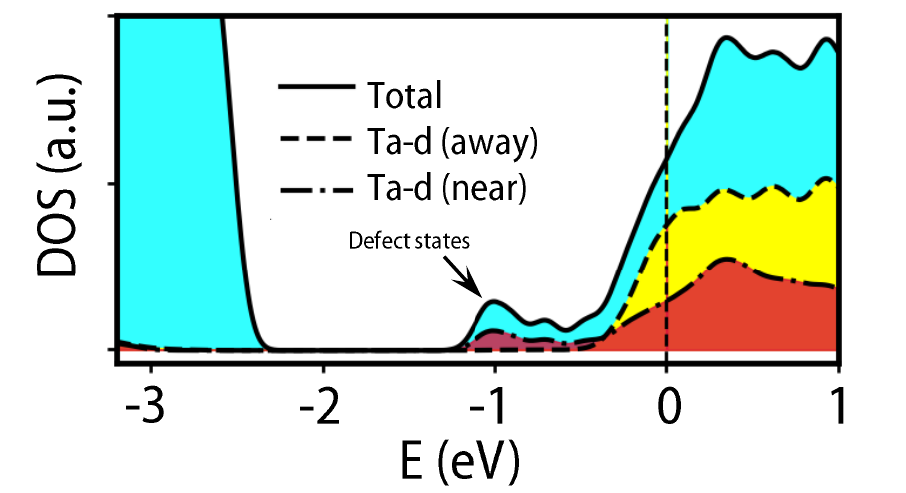}
		\caption{\label{FigS11} DOS plot of oxygen deficient KTO for a supercell of size 2$\times$2$\times$2 with a single OV. }
	\end{figure}
	
	\clearpage

	{\bf \large N. \hspace{0.4 cm} Origin of Skew scattering above 35 K:}
	
	In absence of Rashba SOC,  moments inside an individual BMP becomes collinear along the field direction. Spin dependent asymmetric scattering of itinerant electrons with these giant moments gives rise to observed AHE above 35 K.
	
	\begin{figure} [h]
		\vspace{-0pt}
		\includegraphics[width=0.6\textwidth] {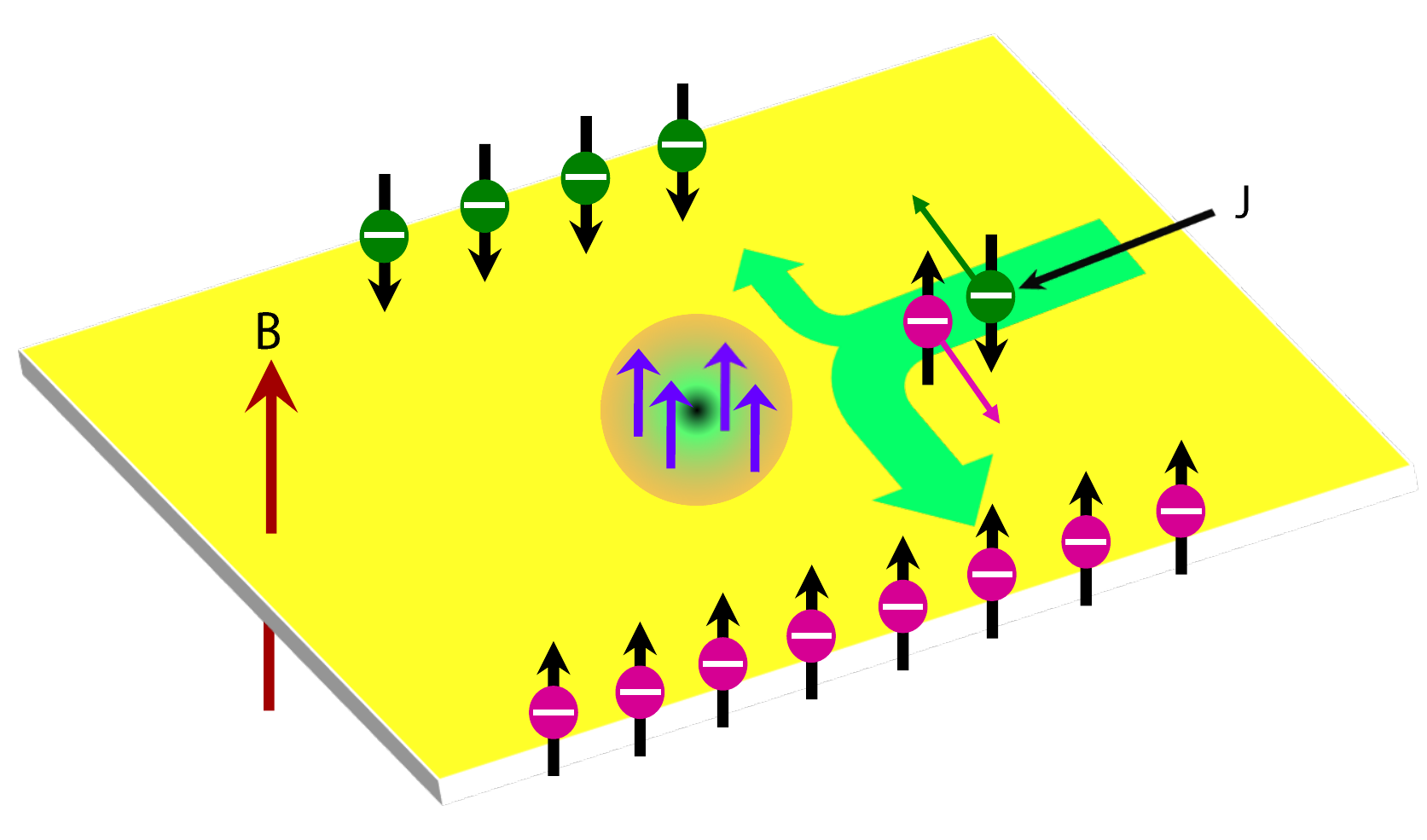}
		\caption{\label{FigS12} Mechanism of skew scattering. }
	\end{figure}

\end{document}